\documentclass[fleqn,usenatbib]{mnras}

\usepackage{newtxtext,newtxmath}

\usepackage[T1]{fontenc}
\usepackage{ae,aecompl}

\usepackage{graphicx}	
\usepackage{amsmath}	
\usepackage{amssymb}	
\usepackage{dcolumn}
\usepackage{bm}
\usepackage{todonotes}

\graphicspath{{./figures/}}

\title[Fingers-of-God Damping in Redshift Space Clustering of LAEs]{Radiative Transfer Distortions of Lyman-\mbox{\boldmath$\alpha$} Emitters:\\ 
a New Fingers-of-God Damping in the Clustering in Redshift Space}

\author[C. Byrohl et al.]{
Chris Byrohl$^{1}$\thanks{E-mail: cbyrohl@mpa-garching.mpg.de},
Shun Saito$^{2,3,1}$,
Christoph Behrens$^{4}$
\\
$^{1}$Max-Planck-Institut f\"{u}r Astrophysik, Karl-Schwarzschild-Stra{\ss}e 1, D-85740 Garching bei M\"{u}nchen, Germany\\
$^{2}$Department of Physics, Missouri University of Science and Technology, 1315 N. Pine Street, Rolla, MO 65409, U.S.A.\\
$^{3}$Kavli Institute for the Physics and Mathematics of the Universe (WPI), 
Todai Institutes for Advanced Study,\\
The University of Tokyo, 5-1-5 Kashiwa-no-Ha, Kashiwa, Chiba 277-8582, Japan\\
$^{4}$Institut f\"{u}r Astrophysik, Georg-August Universit\"{a}t G\"{o}ttingen,
Friedrich-Hund-Platz 1, 37075 G\"{o}ttingen, Germany
}

\date{Accepted XXX. Received YYY; in original form ZZZ}

\pubyear{2019}

\begin{document}
\label{firstpage}
\pagerange{\pageref{firstpage}--\pageref{lastpage}}
\maketitle

\begin{abstract}
Complex radiative transfer (RT) of the Lyman-$\alpha$ photons poses a theoretical challenge to galaxy surveys which infer the large-scale structure with Lyman-$\alpha$ emitters (LAEs). 
Guided by RT simulations, prior studies investigated the impact of RT on the large-scale LAE clustering, and claimed that RT induces a selection effect which results in an anisotropic distortion even in real space but in an otherwise negligible effect in redshift space. 
However, our previous study, which relies on a full radiative transfer code run on the Illustris simulations, shows that the anisotropic selection effect was drastically reduced with higher spatial resolution. 
Adopting the same simulation framework, we further study the impact of RT on the LAE clustering in redshift space. 
Since we measure LAE's radial position through a spectral peak of a Lyman-$\alpha$ emission, the frequency shift due to RT contaminates the redshift measurement and hence the inferred radial position in redshift space. 
We demonstrate that this additional RT offset suppresses the LAE clustering along the line of sight, which can be interpreted as a novel Fingers-of-God (FoG) effect. 
To assess the FoG effect, we develop a theoretical framework with particular emphasis on its connection with the underlying one-point velocity distribution which simultaneously takes into account the RT offset as well as the peculiar velocity that is commonly studied in the context of the Redshift Space Distortion (RSD). 
Although our findings strongly encourage a more careful RSD modeling in LAE surveys, we also seek a method to mitigate the additional FoG effect due to RT by making use of other information in a Lyman-$\alpha$ spectrum.
\end{abstract}




\section{Introduction}
Current and future redshift surveys detecting galaxies with prominent Lyman-$\alpha$ emission, so called Lyman-$\alpha$ emitters (LAEs), can give competitive astrophysical and cosmological constraints.
For example, the Hobby-Eberly Telescope Dark Energy Experiment~\citep[hereafter HETDEX]{Adams2011,Hill2008} is currently operating and will eventually detect close to a million of LAEs in a redshift range of $1.9\leq z\leq 3.5$ over a sky patch of $450\ $deg$^{2}$~\citep{LeungBayesianRedshiftClassification2017}. 
HETDEX will measure cosmic expansion history and growth history of matter fluctuations through baryon acoustic oscillations (BAOs) and redshift space distortion (RSD), and also shed more light on the properties of star forming galaxies by offering a range of statistical measures.\par 

However, HETDEX and other future Lyman-$\alpha$ redshift surveys might
have to deal with severe modifications of the detected spatial clustering signal due to the
complex radiative transfer of Lyman-$\alpha$ given its resonant nature
and high optical depth in astrophysical environments. 
Often Lyman-$\alpha$ photons, particularly those produced 
within the star forming regions of galaxies, scatter many times in the
interstellar medium (ISM) and the circumgalactic medium (CGM) before reaching
the observer. Additionally, column depths of neutral hydrogen outside the 
galaxy's host halo might be sufficient to substantially attenuate the remaining
flux in the intergalactic medium (IGM) by scattering photons out of the
line-of-sight.

Scatterings of neutral hydrogen change both frequency and position of photons
before escaping towards the observer. This can introduce new distortion effects in the 
cosmological signal both in real space and redshift space. In real space this
corresponds to a selection effect favoring certain LAEs to be detected over
others based on their environment, introducing both isotropic and anisotropic
modifications to the two-point statistics. An isotropic distortion
effectively corresponds to changing the bias due to emitters of different mass range being favorably detected. 
Similarly the detection of emitters might be affected
by their large-scale environment, which can also give rise to anisotropic
distortions as demonstrated by~\cite{Zheng2011}.
While these real space distortions automatically propagate into the redshift
space signal, there can be additional distortions purely arising in redshift
space when the radiative transfer modifies the spectral features from which the
line-of-sight position is inferred. Similarly to the selection effects, this can
cause both isotropic and anisotropic distortions in the two-point statistics.

The complexity of the radiative transfer limits analytic solutions to symmetric
toy models, while in physical environments explicit radiative transfer
simulations are needed, which are usually run as a post-processing step given
the expensive numerical cost.
Prior simulations of LAEs in their large-scale structure environment examining
possible distortion effects include \cite{Zheng2010}/\cite{Zheng2011}, 
\cite{BehrensEffectsLymanalpha2013} and~\cite{behrens_impact_2017}.
More studies of LAEs in large cosmological volumes such as
\cite{InoueSILVERRUSHVIsimulation2018} and \cite{Gurung-LopezLyaemitterscosmological2019}
exist, but they employ a more semi-analytic approach assigning spectra to LAEs
based on a range of Monte-Carlo sampled toy models, thus falling short of the
variety of configurations on CGM scales giving rise to the RT interplay on different spatial
scales determining the observed clustering signal.
 
\cite{Zheng2010}/\cite{Zheng2011} find a strong anisotropic selection effect
occurring in the real space clustering signal caused by a correlation of the
observed flux with the large-scale velocity gradient. This selection effect has
been challenged by~\cite{BehrensEffectsLymanalpha2013}
and~\cite{behrens_impact_2017}, who could not reproduce such effect.
\cite{behrens_impact_2017} show that prior findings might have been strongly
overestimated due to a lack of spatial resolution and a simplified emitter
model.

All these studies are primarily concerned with real space distortions, either
not evaluating \citep{BehrensEffectsLymanalpha2013,behrens_impact_2017}
or not finding \citep{Zheng2010,Zheng2011} additional effects in redshift
space. 
In this work, we revisit the idea of possible, additional redshift space distortions of LAEs in redshift space. 
For the analysis, we reuse the radiative
transfer simulations run by~\cite{behrens_impact_2017}, covering a redshift
range from $z=2.0$ to $z=5.85$ as described later.

The structure of this paper is as follows. In Section~\ref{sec:theory}, we describe the
theoretical background needed to model the newly found distortion described
later in the paper. Afterwards, in Section~\ref{sec:methods}, we introduce the
radiative transfer simulations performed by~\cite{behrens_impact_2017} used here
and how to reduce them to mock catalogs of LAEs. In Section~\ref{sec:results},
we present results of detected spectra and
inferred positions before showing how these affect the two-point statistics.
In Section~\ref{sec:discussion}, we summarize our findings along with a discussion of possible
shortcomings in our findings and how future surveys might be affected by/corrected for
radiative transfer redshift space distortions, before concluding in Section~\ref{sec:conclusions}.

\section{Theoretical background of RSD}
\label{sec:theory}
In this section we provide a brief review on the galaxy clustering in redshift space. 
In real space which we considered in our previous work \citep[][]{behrens_impact_2017}, the fluctuation in number density of LAEs for a given sample is given by 
\begin{equation}
    1+\delta_{g}(\vec{x}) = \frac{n_g(\vec{x})}{\bar{n_g}}. 
\end{equation}
Then we consider the two-point statistics to characterize strength of the clustering signal, i.e., the correlation function, $\xi_{g}$, or the power spectrum, $P_{g}$, as its Fourier-counterpart mapped by the Fourier transform (FT)\footnote{
We adopt the following convention for the FT:
\begin{eqnarray}
A(\vec{k}) & = & \int d^{3}x\, e^{i\vec{k}\cdot\vec{x}}A(\vec{x}),\\
A(\vec{x}) & = & \int \frac{d^{3}k}{(2\pi)^{3}}\, e^{-i\vec{k}\cdot\vec{x}}A(\vec{k}).
\end{eqnarray}
}:
\begin{eqnarray}
    1+\xi_{g}(\vec{r}) = \langle\left\{
    1+\delta_{g}(\vec{x})\right\}\left\{
    1+\delta_{g}(\vec{x}+\vec{r})
    \right\}\rangle,\\
    (2\pi)^{3}P_{g}(\vec{k})\delta_{D}(\vec{k}+\vec{k}')
    = \langle \delta_{g}(\vec{k})\delta_{g}(\vec{k}')\rangle,    
\end{eqnarray}
where statistical homogeneity is implicitly assumed. 
In real space where statistical isotropy is given if no selection effect is present, the arguments of the two-point statistics depend only on the scale, i.e., $\xi_g(r)$ and $P_g(k)$.\par 

Now let us consider a contamination of \textit{any form} of velocity, $\vec{v}$, along a line of sight (LOS) of a galaxy in measuring its redshift and hence its radial position. 
Mathematically this is equivalent to a mapping from real to redshift space:
\begin{equation}
    \vec{s} = \vec{r} + \frac{\vec{v}\cdot \hat{r}}{aH(a)}\hat{r}, 
    \label{eq:r2s}
\end{equation}
where $\hat{r}$ is a unit vector along the LOS direction, $a$ is the scale
factor of the Universe, and $H(a)$ is the Hubble expansion rate.
It has been extensively discussed in the literature that the peculiar velocity of a
galaxy contaminates its redshift space position, making the clustering pattern anisotropic known
as RSD \citep[see e.g.,][for a review]{Hamilton:1998}.
This occurs simply because this is the effect only along the LOS, which breaks statistical isotropy.

The complexity to accurately model the two-point statistics in redshift space
originates from the fact that the Jacobian of the mapping in Eq.~\ref{eq:r2s} is
\textit{nonlinear} in terms of the velocity, $v$.
In the following, let us discuss the impact of the nonlinear mapping on the
nonlinear power spectrum and the correlation function as generally as possible.
For this purpose, we begin with the following expression of the redshift-space
density field \citep[][]{TaruyaBaryonAcousticOscillations2010} which is
\textit{exact} under the global plain-parallel approximation \citep[i.e., the
LOS is fixed with one global direction as $\hat{r}\approx \hat{z}$, see][]{Beutler:2014aa}:

\begin{equation}
    \delta^{s}_{g}(\vec{k})=\int d^{3}x\left\{
    \delta_{g}(\vec{x}) - f\partial_{z}u_{z}(\vec{x})
    \right\}e^{i\vec{k}\cdot\vec{x}+ifk_{z}u_{z}(\vec{x})}, 
    \label{eq:deltaS}
\end{equation}
where we introduce a scaled velocity, $\vec{u}\equiv \vec{v}/(faH)$, and $f\equiv d\ln D/d\ln a$ is the linear growth function. 
We specifically denote a quantity in redshift space with a superscript `$s$' throughout this paper. 
We then find an expression for the redshift-space power spectrum as 
\begin{eqnarray}
    P^{s}_{g}(\vec{k}) &=& \int 
    d^{3}r\,e^{i\vec{k}\cdot\vec{r}}\left\langle{e^{-ifk_{z}\Delta u_{z}}}\right.\nonumber\\
    && \left.\times\left\{\delta_{g}(\vec{x}) + f\partial_{z}u_{z}(\vec{x})\right\}
    \left\{\delta_{g}(\vec{x}') + f\partial_{z}u_{z}(\vec{x}')\right\}
    \right\rangle, 
    \label{eq:PS}
\end{eqnarray}
where $\vec{r}\equiv \vec{x}-\vec{x}'$ and $\Delta \vec{u}\equiv\vec{u}(\vec{x})-\vec{u}(\vec{x'})$. 
Eq.~\ref{eq:PS} apparently involves higher-order  correlations between the density $\delta_{g}$ and the velocity field $u_{z}$.  
Linearizing Eq.~(\ref{eq:PS}) in terms of $\delta_{g}$ and $u_{z}$ yields 
\begin{equation}
    P^{s, L}_{g}(\vec{k}) = (1+f\mu^{2})^{2}P_{g}(k), 
\end{equation}
where $\mu$ is the cosine of an angle between $\vec{k}$ and the LOS, defined as $\mu \equiv k_{z}/k$. 
This equation, well known as the Kaiser formula \citep[][]{KaiserClusteringrealspace1987}, shows that the clustering in redshift space is more enhanced closer to the LOS direction, which is a valid picture on large scales and nothing but the main target of RSD measurements \citep[see e.g.,][]{White:2009aa}. 
On the other hand, Eqs.~(\ref{eq:deltaS}) and (\ref{eq:PS}) imply that even random velocity suppresses the redshift-space clustering on small scales along the LOS, often quoted as the Finger-of-God (FoG) effect \citep[][]{JacksoncritiqueReestheory1972}. 
To see this more explicitly, let us rewrite Eq.~(\ref{eq:PS}) in terms of the cumulants as \citep[][]{ScoccimarroRedshiftSpaceDistortionsPairwise2004,TaruyaBaryonAcousticOscillations2010} 
\begin{eqnarray}
 P^{s}_{g}(\vec{k}) &=&\int d^{3}r\,e^{i\vec{k}\cdot\vec{r}}
    \exp\left\{\left\langle e^{-ifk_{z}\Delta u_{z}}\right\rangle_{c}\right\}\nonumber\\
    && \times\left\{ 
    \left\langle e^{-ifk_{z}\Delta u_{z}}\mathcal{A}(\vec{x})\mathcal{A}(\vec{x}')\right\rangle_{c}\right.\nonumber\\
    && \left.\;\;\;\;+\left\langle e^{-ifk_{z}\Delta u_{z}}\mathcal{A}(\vec{x})\right\rangle_{c}
    \left\langle e^{-ifk_{z}\Delta u_{z}}\mathcal{A}(\vec{x}')\right\rangle_{c}
    \right\}, 
    \label{eq:PS_cumulant}
\end{eqnarray}
where $\mathcal{A}(\vec{x})\equiv \delta_{g}(\vec{x}) + f\partial_{z}u_{z}(\vec{x})$ is used just to simplify the notation. 
As \citet[][]{Zheng:2016aa} pointed out, the overall exponential factor, $\exp\left\{\left\langle e^{-ifk_{z}\Delta u_{z}}\right\rangle_{c}\right\}$, contains terms which depend only on the one-point cumulants. 
These terms survive even when two-point correlations such as $\langle \Delta u_{z}\mathcal{A}\rangle$ are zero, and can be integrated out because it no longer depends on the scale. 
That is to say, 
\begin{eqnarray}
 &&\exp\left\{\left\langle e^{-ifk_{z}\Delta u_{z}}\right\rangle_{c}\right\}= \exp\left\{
   \sum^{\infty}_{n=1}(-ifk_{z})^{n}\frac{\left\langle \Delta u_{z}^{n}\right\rangle_{c}}{n!}
   \right\}\nonumber\\
   &=& \exp\left\{
   \sum^{\infty}_{m=1}(-ifk_{z})^{2m}\frac{2\left\langle  u_{z}(\vec{x})^{2m}\right\rangle_{c}}{(2m)!}
   \right\}\nonumber\\
   && \times 
   \exp\left\{
   \sum^{\infty}_{m=1}(-ifk_{z})^{2m}\frac{\left\langle  \Delta u_{z}^{2m}\right\rangle_{c}-\left\langle  u_{z}(\vec{x})^{2m}\right\rangle_{c}-\left\langle  u_{z}(\vec{x}')^{2m}\right\rangle_{c}}{(2m)!}
   \right\}.
   \label{eq:exponential}
\end{eqnarray}
We have used the fact that the terms with odd power in the second and third lines vanish because of symmetry in a galaxy pair. 
The exponential factor in the second line of Eq.~(\ref{eq:exponential}) does not depend on the scale. 
For example, if $u_{z}$ follows a Gaussian distribution with zero mean and variance of $\sigma_{u}^{2}$, only $m=1$ term survives in the exponential factor, corresponding to the FoG damping factor commonly assumed:
\begin{equation}
    \label{eq:DFoG_Gaussian}
    D^{\rm Gaussian}_{\rm FoG}(k, \mu) = e^{-f^{2}k^{2}\mu^{2}\sigma_{u}^{2}}. 
\end{equation}
Notice that, since the FoG damping factor depends only on the one-point cumulants, it can be derived at the level of the density field, Eq.~(\ref{eq:deltaS}). 
Namely, if the velocity field follows a probability distribution function (PDF), $P(u_{z})$, we have 
\begin{equation}
    \label{eq:FoG_FT}
    D_{\rm FoG}(k, \mu) = 
    \left|
    \left\langle e^{ifk\mu u_{z}} \right\rangle
    \right|^{2}
    = \left|
    \int du_{z}\,P(u_{z})e^{ifk\mu u_{z}}
    \right|^{2}
\end{equation}
that is the FT of the one-point PDF, $P(u_{z})$ \citep[e.g.,][]{Hikage:2016aa}. 
Another common velocity PDF is an exponential distribution~\citep[e.g.,][]{ScoccimarroRedshiftSpaceDistortionsPairwise2004}.
The FT of such distribution of the $\exp(-\sqrt{2}\left|u_{z}\right|/\sigma_{u})/(\sqrt{2}\sigma_{u})$, is a Lorentzian damping function\footnote{
We note that the exponential PDF is often adopted for the \textit{pairwise} velocity PDF (see Eq.~(\ref{eq:xiS_pwPDF})) rather than for the velocity PDF \citep[][]{Davis:1983aa,Ballinger:1996aa}. 
In this case, there is no square factor in the damping factor in Eq.~(\ref{eq:DFoG_Lorentzian}). 
We avoid this choice, because the pairwise velocity PDF is generally scale-dependent and hence its mean and dispersion are not necessarily constants.
}
\begin{equation}
    D^{\rm Lorentzian}_{\rm FoG}(k, \mu) = \left\{\frac{1}{1+f^{2}k^{2}\mu^{2}\sigma_{u}^{2}}\right\}^{2}. 
    \label{eq:DFoG_Lorentzian}
\end{equation}
We note again that the PDFs $P(u_z)$ are assumed for the
one-point (rather than pairwise) velocity distributions. This results in a
different interpretation of the $\sigma$ for the Gaussian and the exponential distribution in the literature~\citep[e.g.][]{ScoccimarroRedshiftSpaceDistortionsPairwise2004}.

Here we stress that the FoG damping inevitably arises as long as the velocity field has a non-zero dispersion and higher-order moments. 
For instance, \citet[][]{Agrawal:2017aa} confirmed the damping due to the nonlinear mapping even assuming a linear velocity field.
In addition, the two-point correlations between the density and velocity fields (i.e., the second and third lines in Eq.~(\ref{eq:PS_cumulant})) are essential to accurately model the nonlinear redshift-space power spectrum as several authors have shown~\citep[see e.g.,][]{TaruyaBaryonAcousticOscillations2010,Okumura:2012aa,Matsubara:2014aa,Zheng:2018aa,Vlah:2019aa}.\par

So far we have shown that, in Fourier space, the FoG damping factor depending only on one-point PDF can be expressed as an \textit{overall multiplicative} factor. 
In the following, let us instead discuss the configuration space as a complementary approach. 
The two-point correlation function (TPCF) in redshift space is generally written as \citep[e.g.,][]{ScoccimarroRedshiftSpaceDistortionsPairwise2004}
\begin{equation}
 1+\xi^{s}_{g}(\vec{s}) =  \int d\pi\,\left\{1+\xi_{g}(r)\right\}\mathcal{P}(u_{z};\vec{r}), 
 \label{eq:xiS_pwPDF}
\end{equation}
where $\pi \equiv s_{z}-u_{z}$ is the vector along the LOS direction, $\hat{z}$ in configuration space. 
$\mathcal{P}(u_{z};\vec{r})$ is the \textit{pairwise} velocity PDF given by the FT of the pairwise velocity generating function ${\cal M}(if\gamma,\vec{r})$:
\begin{eqnarray}
  \mathcal{P}(u_z;\vec{r}) & = & \int\frac{d\gamma}{2\pi}e^{-i\gamma u_{z}}
  \mathcal{M}\left( if\gamma;\vec{r} \right),\label{eq:pairwise velocity PDF}\\ 
  \mathcal{M}(if\gamma;\vec{r}) & = & \frac{ 
\langle \exp
\left(if\gamma\Delta u_z \right)\left[1+\delta_{g}(\vec{x})\right]
\left[1+\delta_{g}(\vec{x}')\right]\rangle}{1+\xi(r)}. 
\label{eq:pairwise velocity generating function}
\end{eqnarray}
An advantage of Eq.~(\ref{eq:xiS_pwPDF}) is that the redshift-space correlation function can be expressed only in terms of quantities in real space. 
However, the complexity arises because of a convolution with the pairwise velocity PDF which is weighted by density fields at two points and hence scale-dependent.  
At linear level, the mean of the pairwise velocity PDF is related to coherent infall motion and hence the Kaise factor, while its dispersion is related to the velocity power spectrum \citep[][]{Fisher:1995aa}. 
Similarly to Fourier space at nonlinear level, however, one has to take into consideration the correlation between the density and velocity fields as well as the contribution from the one-point PDF, as several authors have recently studied \citep[see e.g.,][]{ScoccimarroRedshiftSpaceDistortionsPairwise2004,Reid:2011aa,Uhlemann:2015aa,Bianchi:2016aa}.\par 

As we will explain in detail in the next section, we investigate another velocity offset due to the RT effect in addition to the peculiar velocity of a galaxy. 
We will study the impact of the RT velocity component on the redshift-space clustering, mainly focusing on the FoG damping factor in Fourier space, $D_{\rm FoG}(k,\mu)$, and the pairwise velocity PDF. 

\section{Methods}
\label{sec:methods}
\subsection{Radiative Transfer Simulations}

\begin{figure}
\centering
  \includegraphics[width=1.0\linewidth]{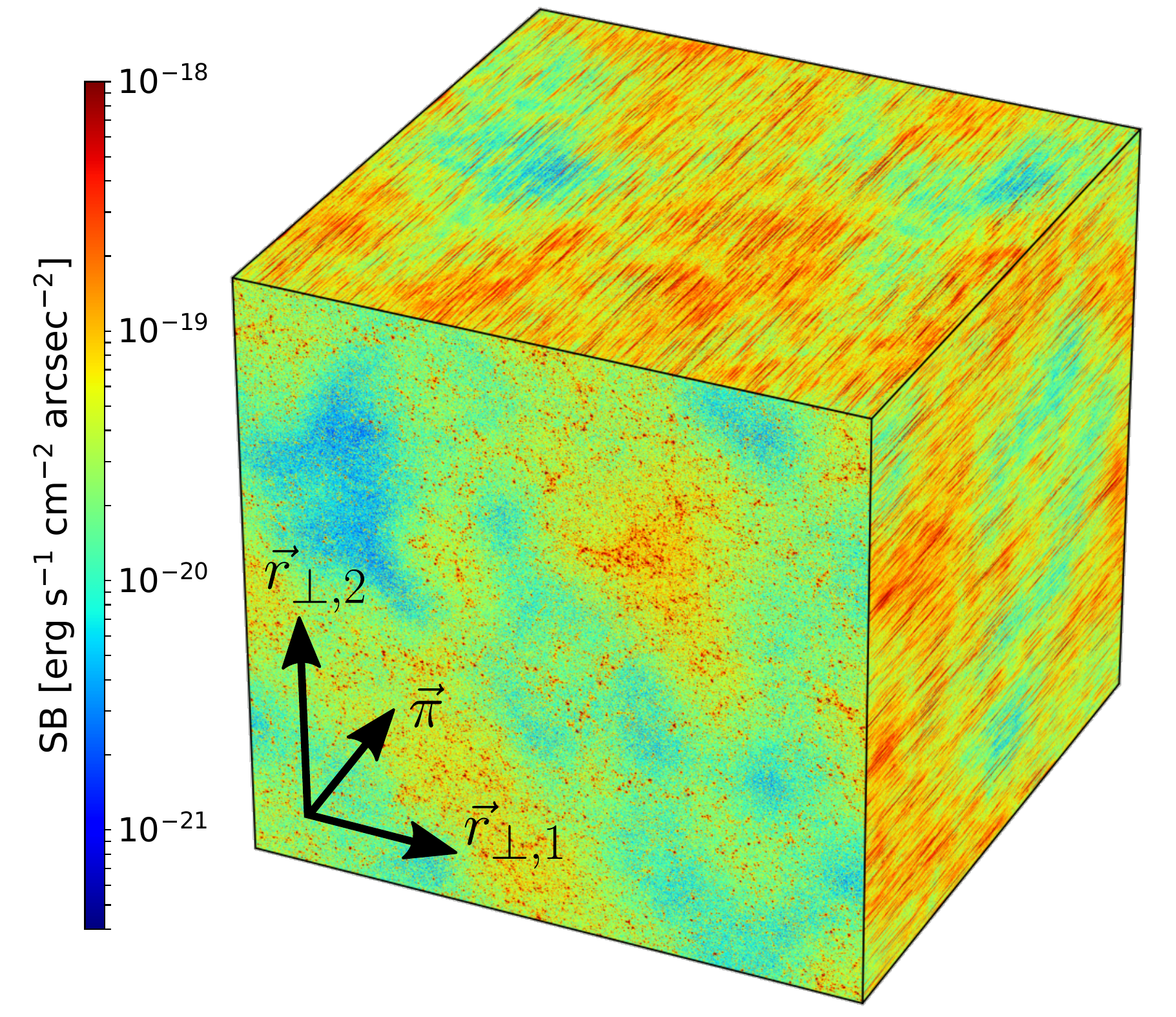}
  \caption{Lyman-$\alpha$ intensity map of the simulation box with a length of
    $75\ $Mpc/h at a redshift of $3.01$ projected onto the cube's faces. $\vec{\pi}$
    denotes the line-of-sight direction, while $\vec{r}_{\perp,1}$ and
    $\vec{r}_{\perp,2}$ indicate the perpendicular direction to former.
    Thus, the top and right cube faces show a directional alignment stemming from
    redshift space distortions. Most of the visible distortion is due to Lyman-$\alpha$
    radiative transfer and the subject of this paper. Individual spectra of LAEs
    are later reconstructed from such cube.}
\label{fig:IMcube}
\end{figure}

We utilize our previous radiative transfer simulations in \citet{behrens_impact_2017} with slight modifications as explained below. 
We run a Monte Carlo radiative transfer code of Ly$\alpha$ photons on top of the Illustris simulations \citep{Vogelsberger2014,Nelson2015} as a post process at redshift outputs of $2.00$, $3.01$, $4.01$ and $5.85$ with its box size of $L_{\rm box}=75\ $cMpc/h.
The Illustris simulations provide a distribution of galaxies and their neutral hydrogen content in a context of the large-scale structure, necessary for studying the impact of Lyman-$\alpha$ radiative transfer onto statistics used in cosmology.
Before applying the radiative transfer, we convert the Voronoi tessellation in Illustris onto an octet-tree data structure with a maximal resolution of $\Delta=3.3\ $ckpc.
The refinement criterion is triggered for cells containing 32 or more Voronoi cells' definining positions. 
More details on the processing of the Illustris datasets can be found in~\cite{behrens_impact_2017}. 

We then explicitly place Ly$\alpha$ photons in a center of the dark matter halos and weight them by the halos' respective Ly$\alpha$ luminosity that is based on star formation rate (SFR) of each halo:
\begin{equation}
\label{eq:lummodel}
L_\textrm{int} = \frac{\mathrm{SFR}}{\mathrm{M}_\odot\,\mathrm{yr^{-1}}} 
\cdot  10^{42}\ \mathrm{erg\,s^{-1}}.
\end{equation}

Note that we assume properties of the LAEs such as positions, velocities and SFR only from host halo catalogs (i.e., `group' catalogs) and hence ignore satellite galaxies. 
We impose a minimum threshold of 0.1 M$_\odot\,\mathrm{yr^{-1}}$ on the SFR and $10^{10}M_\odot$ on the halo mass to limit ourselves to well resolved halos and to limit the required computational resources.
We summarize characteristics of our simulated LAEs in Table~\ref{table1}.

\begin{table}
\caption{In this table, we summarize key quantities of the redshift snapshots considered relevant to our analysis: Spatial resolution $\Delta$, number of LAEs considered
  N$_{\rm LAE}$ and median radius r$_{\rm crit,200}$ encompassing 200 times the critical density of the Universe for those emitters for the post-processed snapshots. Beside the radius' physical size, we also state the angular size as seen for an observer. For each redshift, we also state the average neutral fraction, f$_{\rm IGM}$, at a characteristic hydrogen number density of $10^{-4}$ cm$^{-3}$. We also quote the conversion factor ${(\textrm{aH})}^{-1}$ from peculiar velocity to comoving distance at each redshift.}             
\label{table1}      
\centering          
\begin{tabular}{l | r r r r}  
redshift  & $2.00$ & $3.01$ & $4.01$ & $5.85$ \\
\hline\hline               
f$_{\rm IGM}$ [$10^{-5}$] & $2$ & $3.7$ & $6.8$ & $35$ \\ 
\hline    
$\Delta$ [pkpc] & $1.2$ & $0.8$ & $0.7$ & $0.5$\\
\hline                 
N$_{\rm LAE}$ & $45594$ & $45434$ & $39782$ & $23114$\\
\hline    
r$_{\mathrm{crit,200}}$ [pkpc] & $40.6$ & $30.4$ & $24.3$ & $17.8$ \\
\hline    
${(\textrm{aH})}^{-1}$ [$\frac{\textrm{Mpc h}^{-1}}{\textrm{km s}^{-1}}$] & $0.0105$ & $0.0094$ & $0.0085$& $0.0073$\\
\end{tabular}
\end{table}

Additionally, we set the initial frequency profile emerging from the unresolved
ISM to be a Gaussian whose width $\sigma_i$ is set by the virial temperature of
the halo, see~\citet{behrens_impact_2017}. While the initial profile should have
a significant impact on the observed properties, we lack a profound ISM sub-grid
model for the scope of this paper. For the fiducial sample of LAEs at a number density
of $n_{\rm LAE}=10^{-2}\ $Mpc$^{-3}$h$^{3}$ in our survey, we find a mean input width
$\sigma_i=137\ $km$^{-1}$ at $z=3.01$. This is roughly consistent with recent
findings in shell models by~\cite{GronkeModeling237Lymana2017} finding
$\sigma_i= 172^{+75}_{-60}\ $km s$^{-1}$ fitted to an LAE sample at median
redshift of $z=3.83$.

Varying the Gaussian widths for $T<T_\mathrm{vir}$, we found that this only has
an insignificant impact on the spectra emerging after reprocessing on CGM
scales.
As cosmological simulations such as Illustris are unable to resolve the
ISM regions, we explicitly cut out the unresolved ISM as defined by a hydrogen
number density threshold of $0.13\ $cm$^{-3}$ for the gas. Also, we ignore the
impact of dust attenuation on the radiative transfer for simplicity. Given these
simplifications, we do not expect our simulations to agree well with the
observed luminosity function as already discussed in
\citet{behrens_impact_2017}.

After the luminosity weighted photons are spawned with an isotropic angular distribution in the LAE's rest frame, the photons are propagated in a straight line until a scattering with a neutral hydrogen atom occurs. 
Then at each scattering point, the attenuated contribution along the line-of-sight towards the observer is computed (`peeling-off' photon) while the original photon is re-emitted and propagated/scattered subsequently.

In comparison to the simulations presented in~\cite{behrens_impact_2017}, the only modification in our RT simulation stems from an increased initial Monte Carlo photon count: we increase this count from $100$ to $1000$ in order to properly sample the spectra as a function of wavelength. 
The requirement for the photon count was less important before as only the total flux was relevant for the analysis in~\citet{behrens_impact_2017}. 

As a result of those RT simulations, catalogs of attenuated photon contributions
reaching the observer are created, including information such as the observed
intensity as a function of wavelength and position perpendicular to the line of
sight and the positions of photons' originating LAE. In Figure~\ref{fig:IMcube},
we visualize the reprocessed Lyman-$\alpha$ photons escaping the simulation box and
project the surface brightness onto the cube's faces. The top and the right face
of the cube contain the line-of-sight direction and are shown in redshift space.
One can easily notice a strong anisotropy in redshift space, which will be the
focus of this study. The position of the individually observed LAE along the
line of sight however depends upon a detection algorithm whose methodology we
introduce in the next subsection.

\subsection{Analysis of Simulated LAEs in Redshift Space}

\subsubsection{LAE Spectra and Redshift Space Positions}
\label{sec:methods_catalog}
To determine the position of LAEs in redshift space, we
calculate the flux and spectral information by applying a spherical aperture of 3
arcseconds radius (our default case) around a known LAE's position (from the
halo catalogs), which also already fixes the angular position of the selected
LAE. The aperture size is chosen to correspond roughly to the size of the host halos
in our sample ($4.8$-$3.0$'' for redshifts 2.0 to 5.85; see Table~\ref{table1}). We consider contributions only from the targeted
source, allowing us to separate out any issues due to source confusion (unlike a
real observation). We stress that this detection algorithm is different from the one
in our previous work in real space~\citep[]{behrens_impact_2017} where we
adopted an adaptive Friends-of-Friends (FoF) grouping only along the directions
perpendicular to the line of sight. The previous FoF algorithm naturally leads
to the source confusion, making an interpretation of the redshift-space
clustering more complicated.

The spectrum of a LAE is computed with respect to the comoving frame
at the line-of-sight distance of a given emitter.
We impose a fixed spectral resolution in terms of the velocity shift as
$24.7\ $km/s ($R\sim 12000$) for which we found the Monte Carlo sampling shot
noise to be negligible but still have a sufficient spatial resolution mapping
the velocity differences into comoving distances in redshift space. 
As we show later, the resulting spectra $I_\lambda$ will have one or more peaks.
In this paper, we consider two localization methods for the line-of-sight
position,  $s_\mathrm{app}$, of the LAEs:
Either by using the global spectral maximum at $\lambda_\mathrm{max}$ or by the
spectral maximum at $\lambda_\mathrm{max,red}$ only considering the red wing
with respect to the LAE's rest frame, i.e., $\Delta\lambda > 0$. 

Once a peak has been identified, this allows to define a corresponding apparent
line-of-sight velocity of the emitter as 
\begin{align}
v_\mathrm{app} = c \cdot \frac{\lambda_\mathrm{max}-\lambda_{\mathrm{Ly}\alpha}}
{\lambda_{\mathrm{Ly}\alpha}},
\end{align}
where $c$ is the speed of light, $\lambda_{{\rm Ly}\alpha}$ the rest frame
Lyman-$\alpha$ line center wavelength ($\sim$1216\AA). The comoving position in redshift space is given by Eq.~(\ref{eq:r2s}), i.e., 
\begin{align}
\label{eq:redshiftspace}
\vec{s}_\mathrm{app} = \vec{r}
+\frac{v_{\mathrm{app}}}{aH}\hat{r}.
\end{align}


The apparent line-of-sight position of the emitter is set by adding the apparent
velocity to the position from the halo catalogs.
This completes the localization of the emitters and we construct mock
LAE catalogs containing the position of detected emitters in redshift space as
well as their velocity and spectra.

While we focus on the distortions to the line-of-sight component of the position, we also tested the distortion that RT induces for the angular component. We found that, except for source confusion, deviations from the LAEs' actual
positions to be negligible in comparison to those arising in the line-of-sight direction.

Wavelengths and frequencies in this paper are always evaluated at the emitters' cosmological redshift.
It is important to note that there are two distinct physical velocities contributing to $v_\mathrm{app}$: the peculiar velocity of a halo, $v_\mathrm{pec}$, and the velocity induced by the radiative transfer, $v_{\mathrm{RT}}$. 
In other words, we can estimate $v_{\mathrm{RT}}$ from measured $v_\mathrm{app}$ and $v_\mathrm{pec}$ from the halo catalogs as
\begin{align}
  \label{eq:decomposition}
	v_{\mathrm{app}}=\vec{v}_{\mathrm{pec}}\cdot\hat{r}+v_{\mathrm{RT}}
\end{align}
Velocities and positions without vector notation are the magnitude along the 
line-of-sight if not stated otherwise.
The same decomposition is also done for $\lambda_\mathrm{max,red}$, so that the
radiative velocity is inferred from the peak in the red part of the line center
with respect to the halo's rest frame.

While we choose the host halos' velocity for the peculiar velocity, we checked
that the qualitative reasoning remains the same for other choices such as the
linear velocity field or the velocity of the star-forming regions.
We will show the convenience and significance of this velocity decomposition to
determine $v_{\mathrm{RT}}$ in Section~\ref{sec:results_distinguish_vrt} and~\ref{sec:results_PDF_vrt}.

\subsubsection{Measuring the two-point clustering statistics}

Next, we use the LAE catalogs created in Section~\ref{sec:methods_catalog} to compute the two point statistics, $\xi_{g}(\vec{r})$ and $P_{g}(\vec{k})$.
    
We use \texttt{halotools}
\citep{HearinHighPrecisionForwardModeling2016}\footnote{http://halotools.readthedocs.io/en/latest/}
to compute the TPCF of our LAE samples with the Landy-Szalay estimator
\citep{LandyBiasvarianceangular1993}
\begin{align}
  \label{eq:LSestimator}
  \xi_{g}(\vec{r}) = \frac{\textrm{DD}(\vec{r})
  -2\textrm{DR}(\vec{r})+RR(\vec{r})}{\textrm{RR}(\vec{r})},
\end{align}
where $\vec{r}=\vec{x}-\vec{x}'$ denotes the separation between a pair of
emitters. DD, DR and RR represent LAE-LAE, LAE-random and random-random pair
counts found at the given separation for a given spatial binning width $\Delta
r=\frac{5}{6}\ $Mpc/h.
The pair separation can either be evaluated in real space or redshift space. 
In real space, we expect an isotropic clustering when ignoring RT, so that for such case we can characterize
$\xi(\vec{r})$ as $\xi(r)$ with $r\equiv\left| \vec{r} \right|$. 
As the line-of-sight positions change in redshift space, we express the signal as a function of parallel ($\pi$) and perpendicular ($r_\parallel$) separation to the line-of-sight component.

To estimate $P_{g}(\vec{k})$, we make use of the Fast Fourier Transformation (FFT). 
For this purpose, we first assign LAEs to a three-dimensional grid with 512
cells in each direction (i.e., the Nyquist frequency is $k_{\rm Nyq}\sim 21.4\ $
h/Mpc.) with the Triangle Shape Cloud interpolation.
Next we perform the FFT to obtain the LAE number density field on the grid in
Fourier space, and then measure the power spectrum as
\begin{equation}
  P_{g}(\vec{k}) = \frac{1}{N_{\rm modes}}\sum_{\vec{k}\,{\rm in}\,\,{\rm bin}}
  \delta_{g}(\vec{k})\delta_{g}(\vec{k})^{*}, 
\end{equation}
where $N_{\rm modes}$ is the number of Fourier modes on the grid which fit
within a range of a given bin, e.g., $[k-\Delta k/2, k+\Delta k/2]$ and
$[\mu-\Delta \mu/2, \mu+\Delta \mu/2]$.
We suppress the aliasing effect by applying the interlacing technique as
presented in \citet[][]{Sefusatti:2016ax} and also subtract out the Poisson shot
noise which properly takes the interlacing correction into account. 
prep.).

Another two-point statistics is the pairwise velocity distribution $\mathcal{P}(u_{z},\vec{r})$, which encodes all the information in real space required
to describe induced RSD (see Eq.~\ref{eq:xiS_pwPDF}) on the clustering signal.
We compute the distribution by
\begin{align}
 \mathcal{P}(u_{z},\vec{r}) = \frac{\textrm{DD}(u_{z},\vec{r})}{\textrm{DD}(\vec{r})},
\end{align}
i.e. counting the direct LAE-pairs within at a given separation $r\in\left[r-\Delta r/2, r+\Delta r/2\right]$ and line-of-sight velocity $u_z=\hat{r}\cdot (\vec{v}_2-\vec{v}_1)\in\left[u_z-\Delta u_z/2, u_z+\Delta u_z/2\right]$. The sign convention is chosen such that an infalling motion corresponds to a positive pairwise velocity (also across periodic boundaries). 
For the binning, we chose $\Delta r=1.0\ $Mpc/h and $\Delta u_z=60\ $km/s,
because the number of pairs for central galaxies of interest here quickly goes down below these scales.

\section{Results}
\label{sec:results}
\subsection{Spectra}
\begin{figure*}
\centering
  \includegraphics[width=1.0\hsize]{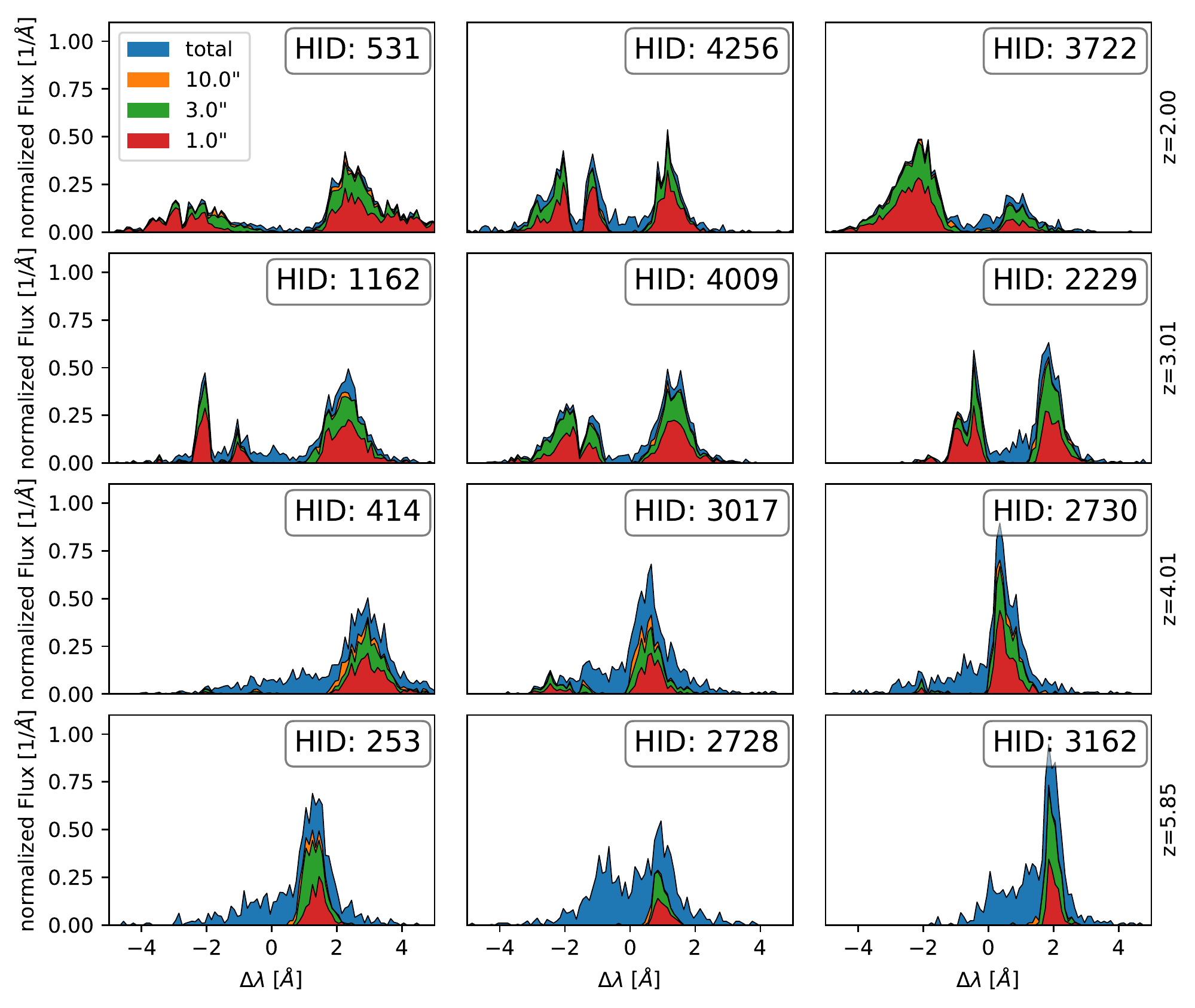}
  \caption{Randomly selected spectra at different redshifts in the comoving
  rest frame at the emitters' position. The halo IDs (HIDs) are given with
  respect to position in Illustris' halo catalogs for given redshift.
  Different colors show the flux for different aperture sizes. As larger
  apertures always enclose the flux of smaller aperture, only the excess flux
  over flux from smaller apertures is shown. The spectra show plenty of
  different characteristics with double peaks being common for lower redshifts
  and rare at high redshifts. Emission becomes more diffuse at higher redshift
  as apparent from larger flux contributions from larger apertures.}
\label{fig:spectra_selection}
\end{figure*}

In Figure~\ref{fig:spectra_selection} we show a random selection of LAE spectra for the simulated redshift range between $z=2.0$ and $z=5.85$.
The spectra are evaluated with respect to the comoving rest frame at the emitters' position.
Most emitters at redshifts $z=2.00$ and $z=3.01$ show a characteristic double
peaked spectrum, which is expected for optically thick environments as shown in fully homogeneous and isotropic analytic toy models \citep{AdamsEscapeResonanceLineRadiation1972,
  Harringtonscatteringresonancelineradiation1973,Neufeldtransferresonancelineradiation1990}
or more recent simulated isotropic shell models for small offset velocities
\citep{AhnCygnitypeLya2003,Verhamme3DLyaradiation2008}.

At $z=4.01$ double peaked spectra become significantly sparser and are only exceptional cases at $z=5.85$. 
This redshift evolution is mostly related to the decreasing transmissivity in the IGM at higher redshifts \citep{laursen_intergalactic_2011} due to the increasing neutral hydrogen density, explaining the disappearing of blue peaks at high redshift and a stronger trough around the line center. 
At the same time the spatial luminosity profiles become more extended as scattering in the CGM increases with redshift. 
This trend shows up in Figure~\ref{fig:spectra_selection}, as relative contributions to the flux 
at fixed finite aperture becomes lower at higher redshift. 

Our simulations seem to overestimate the abundance of double peaked emitters when compared to observations at high redshifts $z=2.0-3.0$, where the IGM
interaction is limited. Typically observations find fractions of $\sim30\%$ at
these redshifts in star-forming galaxies (\cite{KulasKinematicsMultiplePeakedLyalpha2012b}, \cite{HerenzMUSEWidesurveyfirst2017}),
while most of our emitters show double peaked emitters. We shortly discuss the shortcomings in the modeled spectral shape and thus also the overprediction in the double-peaked profiles in Section~\ref{sec:discuss_localization}.\par 

More complex spectra are present depending on the underlying HI distribution and velocity structure, such as emitters with $n\geq3$ peaks particularly in the blue part of the spectrum. See HID 4256 at z=2.0 or HID 1162 at z=3.01 for such examples.

\begin{figure*}
\centering
  \includegraphics[width=1.0\linewidth]{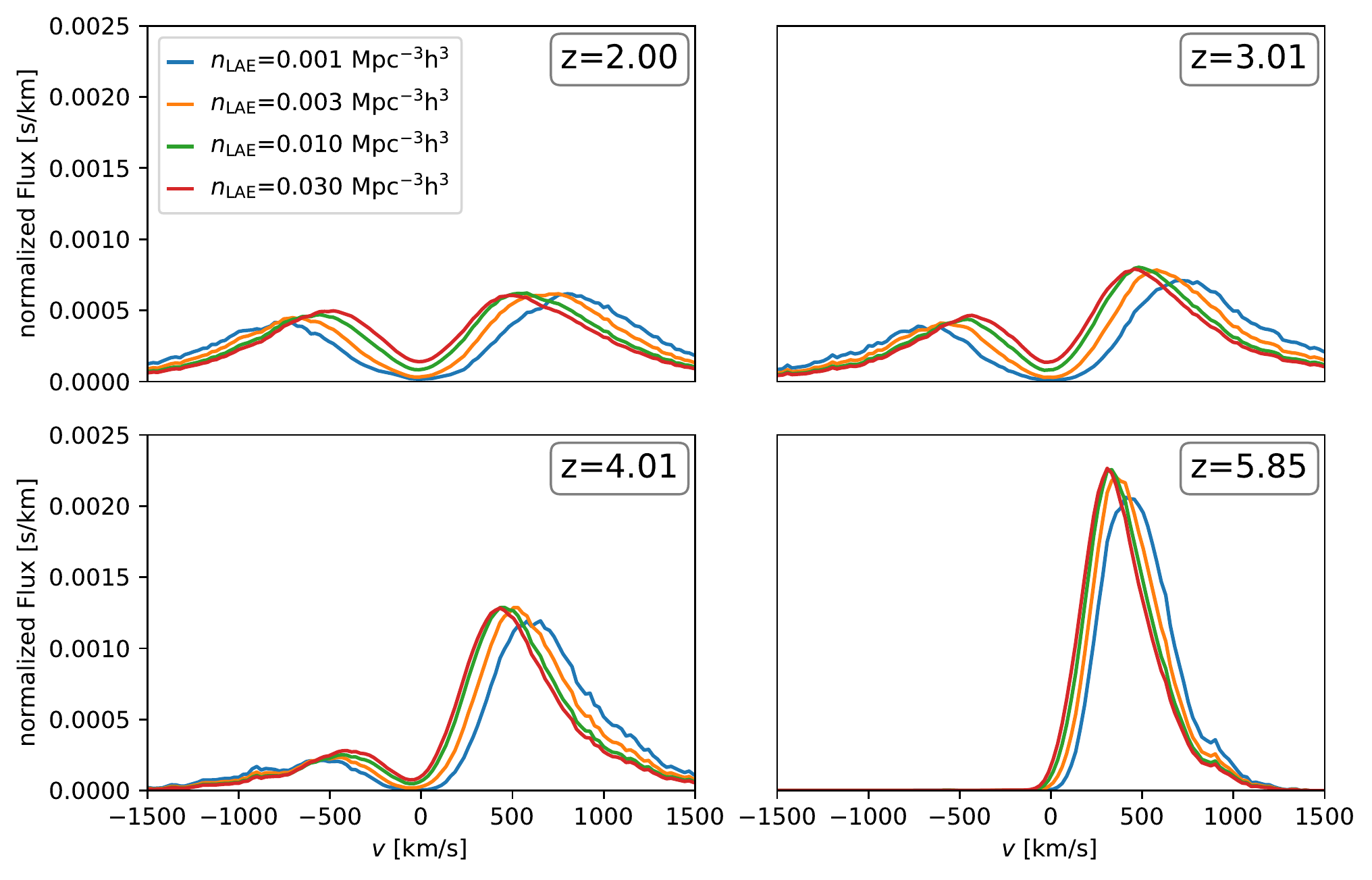}
  \caption{Stacked spectra in the halos' frame for different number
  densities and redshifts for an aperture radius of 3 arcseconds. The overall
  flux is normalized with respect to the sample of chosen number density $n_{\rm
  LAE}$.}
\label{fig:vPDF_stacked}
\end{figure*}

The redshift evolution becomes particularly apparent when computing the stacked profiles $I_{\rm stacked}$
of the emitters as shown in Figure~\ref{fig:vPDF_stacked}. 
Here we vary redshifts and LAE number densities $n_{\rm LAE}$ for which we
consider only the $n_{\rm LAE}L_{\rm box}^{3}$ emitters with the highest
apparent luminosity.
Imposing a number density threshold has several advantages when compared to a
surface brightness threshold, including independence from the proportionality
constant in the luminosity model in equation~\eqref{eq:lummodel} and controlled
shot noise behavior for measuring LAE power spectra.
The spectra are stacked in the halos' rest frame. 
Stacking in the halo's rest frame reveals a trough at $v=0\ $km/s in the spectra
caused by a combination of IGM attenuation and strong frequency diffusion into
the wings due to high optical depths.
As the number density $n_{\rm LAE}$ is reduced, the peaks of the stacked
profiles move to higher offsets. At the same time, the dispersion $\sigma_{\rm
  stacked}$, the stacked profiles' second central moment for the respective red or
blue spectral peak, also slightly increase but only on the
percent level. We motivate the significance of this dispersion in
Section~\ref{sec:discuss_localization}. The change in $\sigma_{\rm stacked}$
mostly stems from a significant correlation between brighter sources and more
massive sources which in turn have a larger trough/peak separation.

\subsection{Distinguishing Distortion Contributions}
\label{sec:results_distinguish_vrt}

\begin{figure}
\centering
  \includegraphics[width=1.0\linewidth]{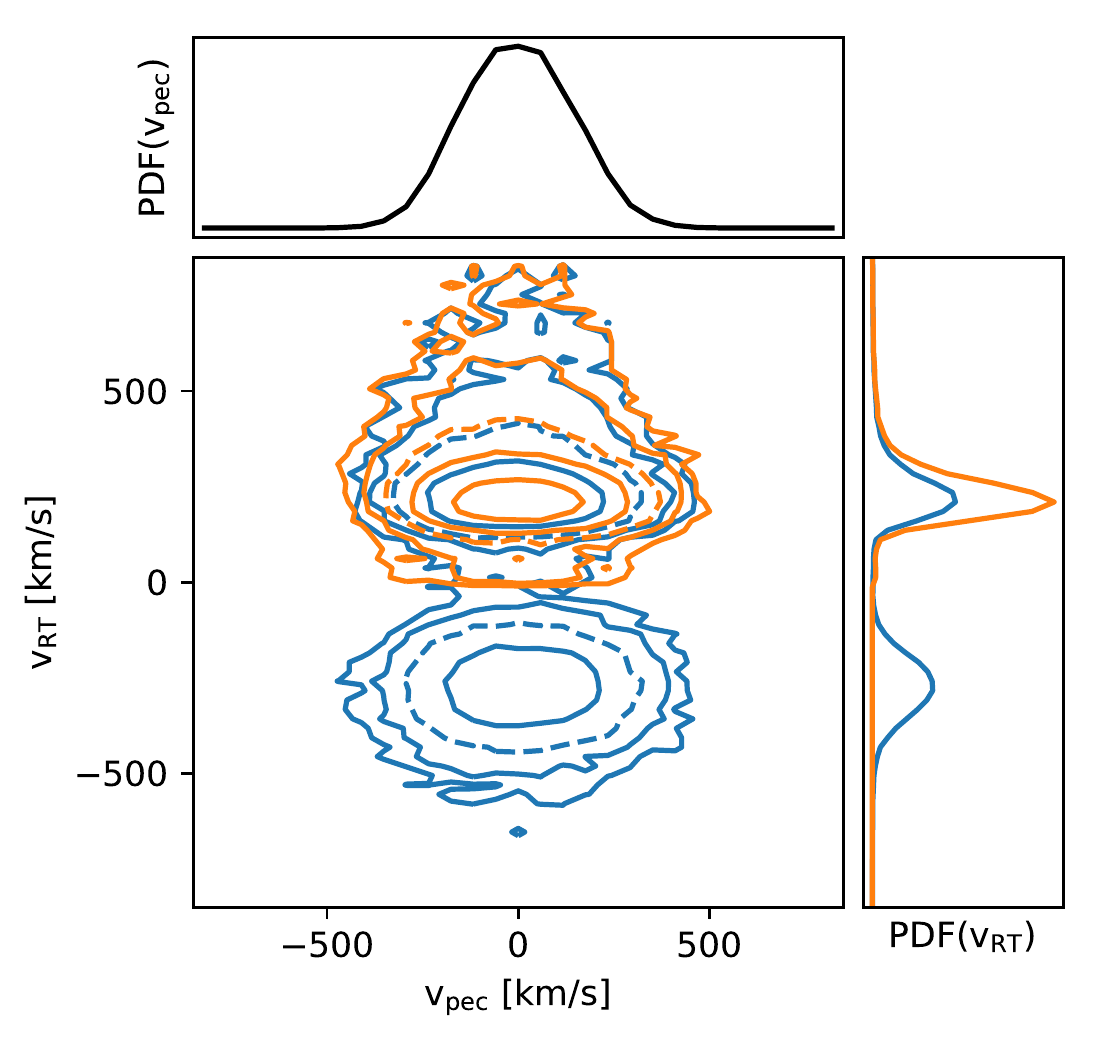}
  \caption{Contour plot of emitters' peculiar and radiative velocities.
	  Detection of the global peak (\textbf{blue}) and the peak in the
	  red wing of the emitters' rest frame (\textbf{orange}).}
\label{fig:vRT_vs_vpeculiar_z3.01}
\end{figure}

Figure~\ref{fig:vRT_vs_vpeculiar_z3.01} shows the one-point PDF, $P(v)$, decomposed into $v_\mathrm{RT}$ and $v_\mathrm{pec}$ for $z=3.01$ as suggested in Eq.~(\ref{eq:decomposition}).
The projected PDFs onto the $v_{\rm pec}$ and $v_{\rm RT}$ axes give the one-point PDF, $P(v_\mathrm{pec})$, and $P(v_\mathrm{RT})$, respectively. 
There is no noticeable correlation between the two velocity components in Figure~\ref{fig:vRT_vs_vpeculiar_z3.01}. 
We checked that the same holds true for other redshift outputs as well. 
This result is expected to some extent, since two physically distinct processes are responsible for each velocity component. 
As we discuss later, this independence allows us to model additional RSD due to the RT effect independently of the peculiar velocity.

\subsection{RT Velocity PDF}
\label{sec:results_PDF_vrt}
\begin{figure*}
\centering
  \includegraphics[width=1.0\linewidth]{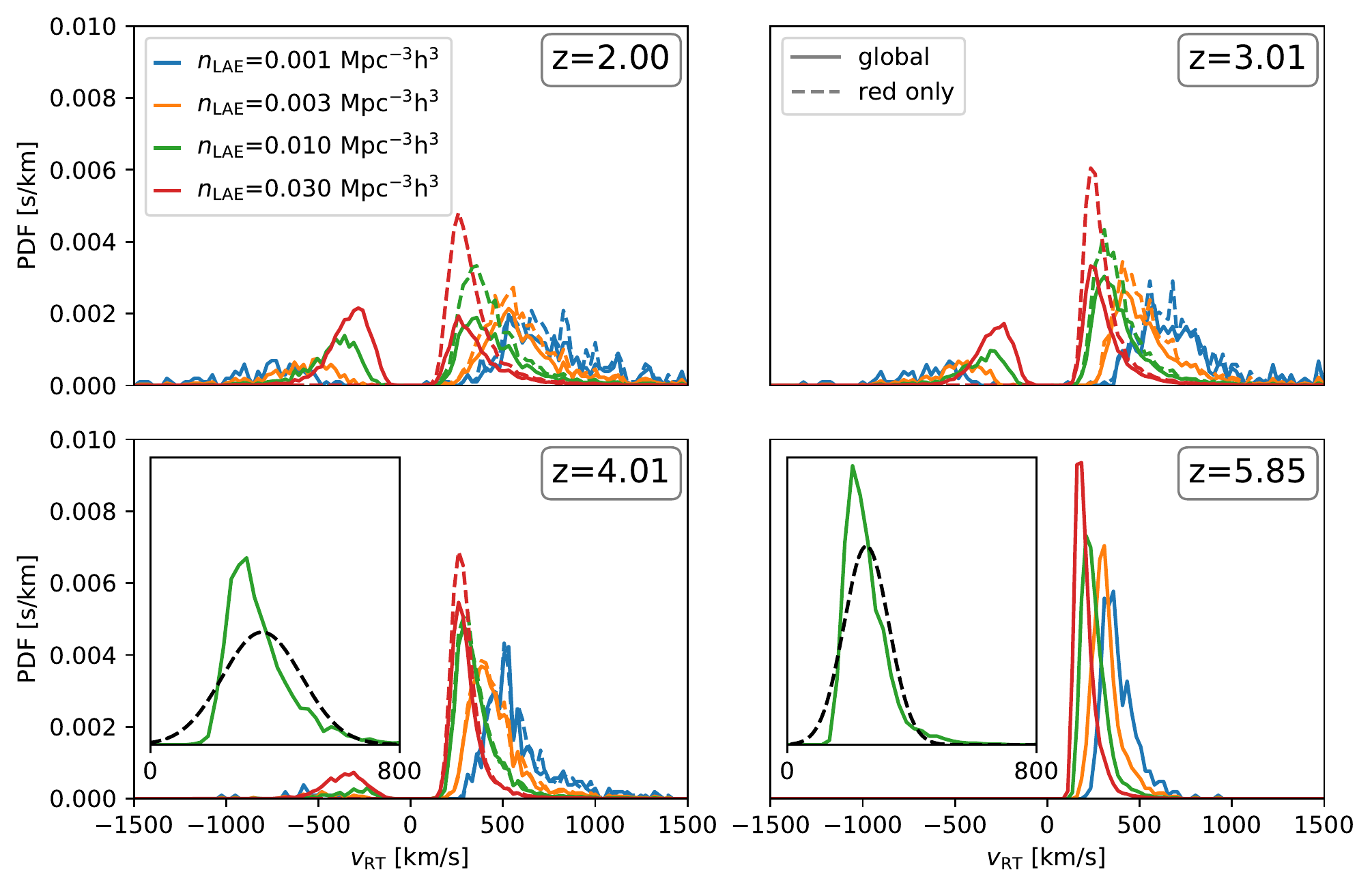}
  \caption{Radiative transfer velocity $v_{\rm RT,global}$/$v_{\rm RT,red}$ distributions for different number
  densities $n_{\rm LAE}$ and redshifts. Solid lines show the velocity being
  deduced from the global peak of each LAE's spectra and dashed lines with the
  velocity deduced from the red peak only. For redshift $z=3.01$ and $4.01$ we
  show the distribution of $v_{\rm RT,global}$ for the red peak at $n_{\rm
    LAE}=0.01\ $Mpc$^{-3}$h$^{3}$ along with a Gaussian of matching mean and standard
  deviation. The scaling of the y-axis remains the same as for the other
  subplots. The velocity distributions for the respective wings are positively
  skewed with respect to the velocity offset to the line center.}
\label{fig:vPDF}
\end{figure*}

As we discussed in Section \ref{sec:theory}, 
the velocity PDF, i.e. a probability distribution of radial position of LAEs
with respect to real-space position (see e.g., Eq.~(\ref{eq:FoG_FT})),
determines the redshift space clustering.
We show the PDF of the peaks for the brightest emitters detected by the
observer for a given LAE number density $n_\mathrm{LAE}$ in
Figure~\ref{fig:vPDF}.
The peak distributions show a strong redshift evolution: As we move to higher redshifts, the blue peaks are strongly suppressed and practically non-existent at z=5.85. 
When decreasing the number density, the selected emitters are restricted to the more luminous ones, which in turn have a higher average peak offset due to their higher optical depth.\par

Although we have already seen similar trends in the stacked
spectra as expected, it is important to notice that the velocity PDFs and the
stacked spectra are \textit{not} the same.
We discuss a possible correlation between the distribution of stacked spectra and the velocity PDF in Section~\ref{sec:discuss_localization}. 
Since only the stacked spectra are directly observable, such relation will prove itself crucial in estimating the additional radiative transfer effect in observational surveys stemming from the velocity PDF.\par

In Figure~\ref{fig:vPDF}, we also provide a comparison of the $v_\mathrm{RT}$-PDF with a Gaussian of same mean and variance for the red peaks. We find that the distributions quickly fall off towards the line center while they extend more towards larger velocity offsets. This leads to a higher kurtosis than for the Gaussian and also a positive skewness, which will affect the quality of using a Gaussian approximation for the damping in the next sections.

\begin{figure*}
\centering
  \includegraphics[width=0.88\hsize]{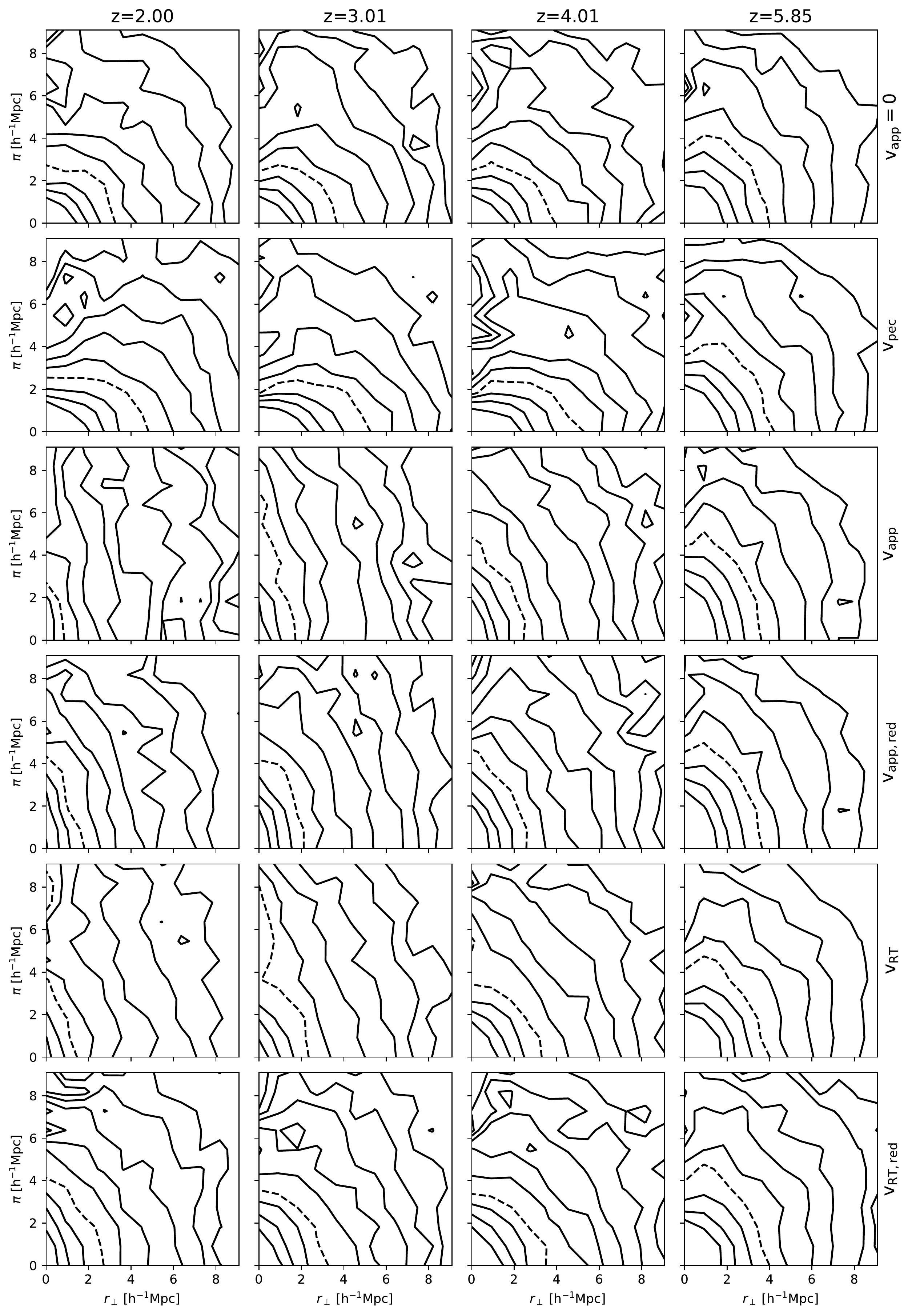}
  \caption{TPCF for disentangled RSD of visible LAEs for $n_{\rm LAE}=10^{-2}$
    h$^{3}$Mpc$^{-3}$ in redshift space. The dashed contour corresponds to
    $\xi=1$ with contours decreasing by a factor of $1.4$ further away from the
    origin. In the first row, we set the apparent velocity $v_{\rm app}$ to
    zero, i.e. we show the real space clustering. In the second row, we only
    consider the peculiar velocity $v_{\rm pec}$ of the corresponding host halo. The third
    and fourth row show the apparent overall velocity shifts $v_{\rm app}$ detected from the
    global peak and the red peak respectively. The last two rows show the clustering when only the
    radiative transfer contribution is considered.}
\label{fig:TPCF_distangled_visemitters}
\end{figure*}

\subsection{Configuration Space: TPCF and the paiwise velocity PDF}
In Figure~\ref{fig:TPCF_distangled_visemitters} we show the correlation functions $\xi(\pi,r_{\parallel})$ measured from the mock observations with a LAE number density of $n_{\rm LAE}=0.01\ $h$^3$/Mpc$^3$. 
Different columns show the clustering at different redshifts in an increasing order, while different rows show different velocity contributions (as defined in Section~\ref{sec:methods_catalog}) added onto the real space configuration to obtain the redshift space clustering.
In the first row we set the velocity contribution $v_{\rm app}$ to zero, so that we plot the real space clustering.
The second row shows the redshift space result using the peculiar velocity $v_{\rm pec}$ from the halo catalogs and thus explicitly omitting the contribution from radiative transfer. 
We stress that the second row is often presented as the redshift-space clustering of LAEs in the literature \citep[][]{Zheng2011,Gurung-Lopez:2019aa} but does not yet directly correspond to an observable radial position from the redshift measurement. 
Instead, the apparent $v_{\rm app}$ is the observable containing both contributions from the complex radiative transfer and the peculiar velocity.
The third/fourth ($v_{\rm app}$/$v_{\rm app,red}$) row shows inferred overall position from the peaks in the spectra, which includes both peculiar velocity and radiative transfer effects.
The fifth/sixth ($v_{\rm RT}$/$v_{\rm RT,red}$) rows show the radiative transfer component of the velocity only as the residual of the apparent and peculiar velocity.

Given the same simulation setup as \cite{behrens_impact_2017}, we expect a very similar result to those shown for the real-space clustering in the first row. 
Only minor differences arise from an increased Monte Carlo photon count and the simplified detection algorithm.
Note that there is a slight anisotropy in the clustering signal in real space.
As we stressed in ~\cite{behrens_impact_2017}, the slight anisotropy of this
dataset does not originate from a radiative transfer effect and was statistically consistent with zero.


As introduced in Section~\ref{sec:theory}, we confirm two competing RSD effects in Figure~\ref{fig:TPCF_distangled_visemitters}.
We see that on the shown scales ($1-10$h$^{-1}$Mpc) the Kaiser effect dominates
the redshift space distortions from the peculiar velocity field $v_\mathrm{pec}$ over the usual FoG effect due to random motion of the LAEs in the second row.
Once $v_{\rm RT}$ is added, however, the small-scale damping from $v_\mathrm{RT}$ is significant on these scales and thus the overall redshift space clustering with the apparent velocities $v_\mathrm{app}$ is elongated along the line-of-sight despite the squashing from the Kaiser effect (third and fourth row).\par 

In order to investigate the distortions of the TPCF in redshift space more quantitatively, we discuss the pairwise velocity PDF which encodes the full information of RSD (see Eq.~(\ref{eq:xiS_pwPDF})). 
Note that we do not report the measurement of the multipole moment as we did in \citet[][]{behrens_impact_2017}.
Before showing the pairwise velocity PDF, let us first extend the discussion in \ref{sec:theory} in the presence of two physically-distinct velocity contaminations $v_{\rm pec}$ and $v_{\rm RT}$ (see Eq.~(\ref{eq:decomposition})). 
In Sec.~\ref{sec:results_distinguish_vrt}, we show that there is no apparent correlation between $v_{\rm pec}$ and $v_{\rm RT}$ at the level of the one-point PDF, i.e., $\langle v_{\rm pec}(\vec{x})v_{\rm RT}(\vec{x})\rangle=0$. 
We further assume that $v_{\rm RT}$ has no correlation with the density field or the peculiar velocity at scales of interest, i.e., $\langle u_{\rm RT}(\vec{x})\delta_{g}(\vec{x}')\rangle=\langle u_{\rm RT}(\vec{x})u_{\rm pec}(\vec{x}')\rangle=0$. 
Under this simple setting, Eq.~(\ref{eq:pairwise velocity generating function}) follows that
\begin{equation}
      \mathcal{M}(if\gamma;\vec{r})  =  
\frac{ 
\langle 
e^{if\gamma\Delta u_{{\rm pec},z}}\left[1+\delta_{g}(\vec{x})\right]
\left[1+\delta_{g}(\vec{x}')\right]\rangle}{1+\xi(r)}
\langle e^{if\gamma\Delta u_{{\rm RT},z}}\rangle. 
\label{eq: total M}
\end{equation}

Its FT, the pairwise velocity PDF, is written as 
\begin{equation}
    \mathcal{P}(u_z;\vec{r}) = (\mathcal{P}_{\rm pec}*\mathcal{P}_{\rm RT})(u_z;\vec{r}), 
    \label{eq: pPDF convolution}
\end{equation}
where $*$ denotes the convolution for simplicity with $\mathcal{P}_{\rm RT}$ given by
\begin{eqnarray}
  \mathcal{P}_{\rm RT}(u_{z}) & = & \int \frac{d\gamma}{2\pi}e^{-i\gamma u_{z}} \langle e^{if\gamma\Delta u_{{\rm RT},z}}\rangle\nonumber\\
   & = &\int \frac{d\gamma}{2\pi}e^{-i\gamma u_{z}}\left|\int du\,P_{\rm  RT}(u)e^{if\gamma u}\right|^{2}.
   \label{eq: pPDF RT}
\end{eqnarray}
Notice that the $u_{\rm RT}$ contribution in Eq.~(\ref{eq: total M}) is not weighted by the density field at different scales and, as a result, the ensemble average becomes an integration over one-point PDF, $P_{\rm  RT}(u_{z})$. 
In other words, the scale dependence in the pairwise velocity PDF in Eq.~(\ref{eq: pPDF convolution}) comes only from the peculiar velocity part, $\mathcal{P}_{\rm pec}(u_{z};\vec{r})$. 

\begin{figure}
\centering
\includegraphics[width=1.0\linewidth]{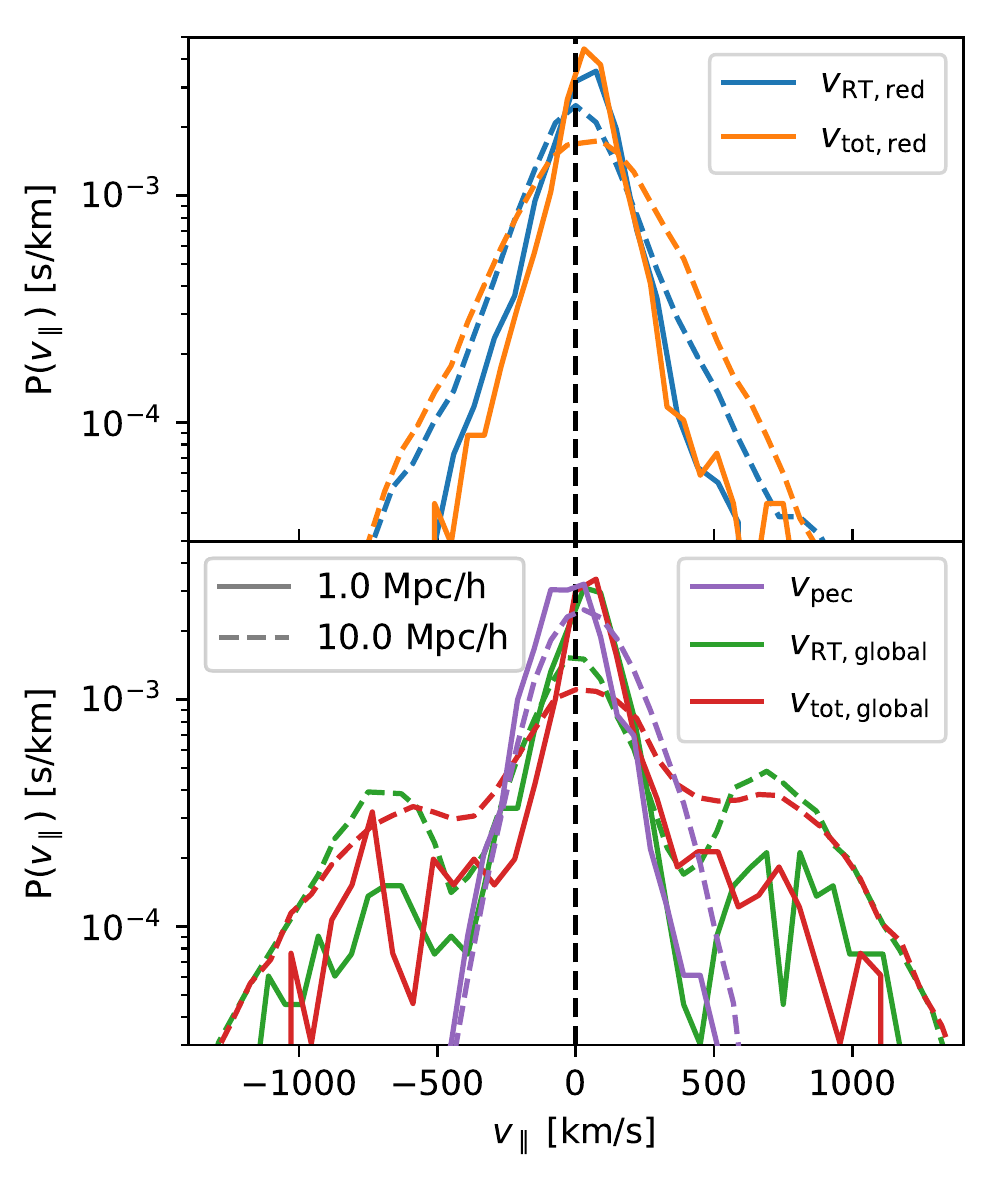}
  \caption{The pairwise velocity distribution ${\cal P}$ for the peculiar and radiative transfer velocity component evaluated at different length scales for $z=3.01$ with $n_{\rm LAE}=0.01\ $h$^3$/Mpc$^3$ along the line-of-sight. The upper panel shows the radiative transfer and total apparent velocities ($v_{\rm RT,red}$/$v_{\rm tot,red}$) based on the red peak only detection method. The lower panel shows the same for the global peak detection method plus the peculiar velocity distribution $v_{\rm pec}$. Different line styles indicate different spatial separations $\vec{r}$.}
\label{fig:vPDF_pairwise_scaledependency_z301}
\end{figure}

In Figure~\ref{fig:vPDF_pairwise_scaledependency_z301}, we show the measured pairwise velocity PDFs at two different scales (solid and dashed lines for 1 cMpc/h and 10 cMpc/h, respectively) in cases of both the red-peak only (upper panel) and the global peak (lower panel) at $z=3.01$. 
First of all, when we adopt $v_{\rm pec}$ (purple lines) as a velocity, we confirm a trend well known in the literature \citep[see e.g.,][]{ScoccimarroRedshiftSpaceDistortionsPairwise2004};
We see a positive peak which corresponds to a coherent infall motion on large scale ($r_{\parallel}=10\ $cMpc/h), while the PDF follows a distribution with an exponential tail at small separation, $r_{\parallel}=1\ $Mpc/h.
On the other hand, we do not confirm such a trend in the case of $v_{\rm RT}$. 
In general, a large tail of the pairwise $v_{\rm RT}$-PDF contributes to that of the pairwise $v_{\rm app}$-PDF even at a large separation of 10 cMpc/h. 
This impact is clearly more significant for the global peak only case than for the red-only peak case.

\begin{figure}
\centering
\includegraphics[width=1.0\linewidth]{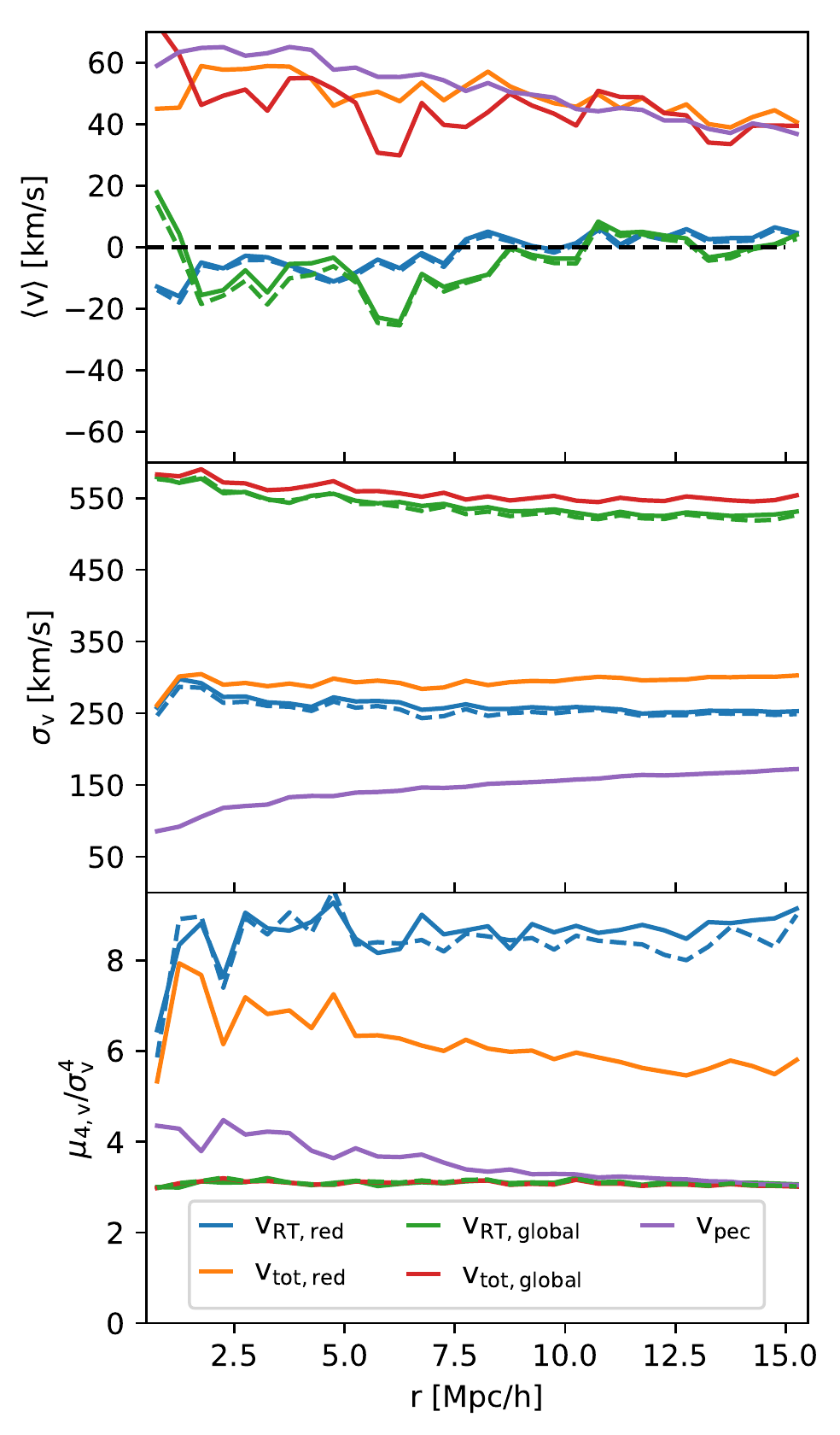}
  \caption{The mean, standard deviation and kurtosis of the pairwise velocity
    distribution ${\cal P}$ for the peculiar and radiative transfer velocity
    component as a function of the total separation $r$ of emitter pairs at $z=3.01$ with $n_{\rm LAE}=0.01\ $h$^3$/Mpc$^3$ along the line-of-sight. The solid lines show the respective velocity contributions as listed in the legend, ie.. the apparent and radative transfer velocity for each the global and red-only detection method along with the peculiar velocity. In the dashed lines we show the expected radiative transfer velocity contribution $v_{\rm RT}$ at a given scale under the assumption that $v_{\rm RT}$ and $v_{\rm pec}$ are independent distributions. Thus, the difference between the solid and dashed line indicates the strength of any correlation between these two contributions present on a given scale. The color coding is chosen to be consistent with Figure~\ref{fig:vPDF_pairwise_scaledependency_z301}.} 
\label{fig:vPDF_pairwise_scaledependency_3in1_z3}
\end{figure}

To see the scale dependence of the pairwise velocity PDFs more explicitly, we show three low-oder moments of the PDFs  (mean, dispersion, and kurtosis) as a function of a separation scale in Figure~\ref{fig:vPDF_pairwise_scaledependency_3in1_z3}, where we further confirm the aforementioned trends.
The mean infall velocity (upper panel) is dominated by $v_{\rm pec}$ particularly on large scales, and the impact of $v_{\rm RT}$ on the mean is marginal even in the global peak case. 
This is the reason why the Kaiser squashing effect is seen on large scales in Figure~\ref{fig:TPCF_distangled_visemitters}. 
Meanwhile, the dispersion (middle panel) is mainly dominated by $v_{\rm RT}$ for both detection algorithms, and notably has little scale dependence. 
Interestingly, the dominant contribution to the kurtosis (lower panel) depends on the peak finding algorithm: 
As expected, the kurtosis of $v_{\rm pec}$ deviates from the Gaussian value of 3 at small scales due to its extended tail \citep[e.g.,][]{ScoccimarroRedshiftSpaceDistortionsPairwise2004}. 
In the case of the red peak detection method, a much wider distribution due to $v_{\rm RT}$ dominantly contributes to the kurtosis than for $v_{\rm pec}$. 
On the other hand, both $v_{\rm pec}$ and $v_{\rm RT}$ roughly equally contribute to the kurtosis for the global peak detection method.  
We do not find a strong scale dependence of the kurtosis for both cases. 
Since the dispersion and higher-order moments of the pairwise velocity PDF contribute to the FoG elongation \citep{ScoccimarroRedshiftSpaceDistortionsPairwise2004}, this evidence suggests that random velocity following the one-point PDF plays a main role in the FoG elongation in Figure~\ref{fig:TPCF_distangled_visemitters}.

In Figure~\ref{fig:vPDF_pairwise_scaledependency_3in1_z3}, we also show the three moments of the $v_{\rm RT}$ estimated from $v_{\rm tot}$ and $v_{\rm pec}$, assuming that $\mathcal{P}_{\rm pec}$ and $\mathcal{P}_{\rm RT}$ are independent of each other (dashed lines compared with solid lines with the same colors). 
Although they are roughly consistent, we see small discrepancies between solid and dashed lines, implying that either $\langle v_{\rm RT}(\vec{x})\delta_{g}(\vec{x}')\rangle$ or $\langle v_{\rm RT}(\vec{x})v_{\rm pec}(\vec{x}')\rangle$ is not exactly equal to zero. 

\subsection{Fourier Space: The FoG damping factor}
\label{sec:damping}

In the previous subsection, we show that the impact of $v_{\rm RT}$ on the pairwise velocity PDF can be roughly understood in terms of the one-point PDF, $P_{\rm RT}(u_{z})$. 
However, it is not straightforward to quantify its impact on the observable TPCF because of the involved convolution (see Eq.~(\ref{eq: pPDF convolution})). 
It is simpler to work in Fourier space, since the convolution becomes a multiplication after FT. 
Since Eq.~(\ref{eq: pPDF convolution}) follows 
\begin{equation}
    \label{eq:xiRT}
    1+\xi^s_{g}(r_\perp,\pi) = \left(\left[ 1+\xi^s_{g,{\rm pec}}(r_\perp,\pi) \right]
    * {\cal P}_{\rm RT}\right) (u_{z}), 
\end{equation}
we obtain the FoG damping in Fourier space with a help from Eq.~(\ref{eq: pPDF RT}),
\begin{equation}
\label{eq:dampingfactor}
    D^\mathrm{RT}_\mathrm{FoG}(k,\mu) \equiv 
\frac{P^{s}_{g,{\rm tot}}(\vec{k})}{P^{s}_{g,{\rm pec}}(\vec{k})}
= \left|\int du_{z}\,P_{\rm  RT}(u_{z})e^{ifk\mu u_{z}}\right|^{2}, 
\end{equation}
where $P^{s}_{g,{\rm tot}}(\vec{k})$ and $P^{s}_{g,{\rm pec}}(\vec{k})$ denote the redshift-space power spectra  
when we adopt the total apparent velocity and the peculiar velocity only as a velocity contamination, respectively. 
This relation allows us to directly compare the line-of-sight damping from radiative transfer as found in the measured redshift-space power spectra with an expected damping from the underlying one-point PDF. 
We stress that the last equality holds only under the assumptions we made in Eqs.~(\ref{eq: total M}) and (\ref{eq: pPDF convolution}): namely, no correlation between $v_{\rm RT}$ and $v_{\rm pec}$ at the same point, and no spatial correlations between $v_{\rm RT}$ and $\delta_{\rm g}$ (and $v_{\rm pec}$).  
Furthermore, we will compare this damping with the two generic functions commonly adopted in the literature, i.e., the Gaussian and the Lorentzian damping functions (see Eqs.~(\ref{eq:DFoG_Gaussian}) and (\ref{eq:DFoG_Lorentzian})). 
Here the second central moment $\sigma$ is directly calculated from the one-point PDF.

\begin{figure*}
\centering
  \includegraphics[width=1.0\linewidth]{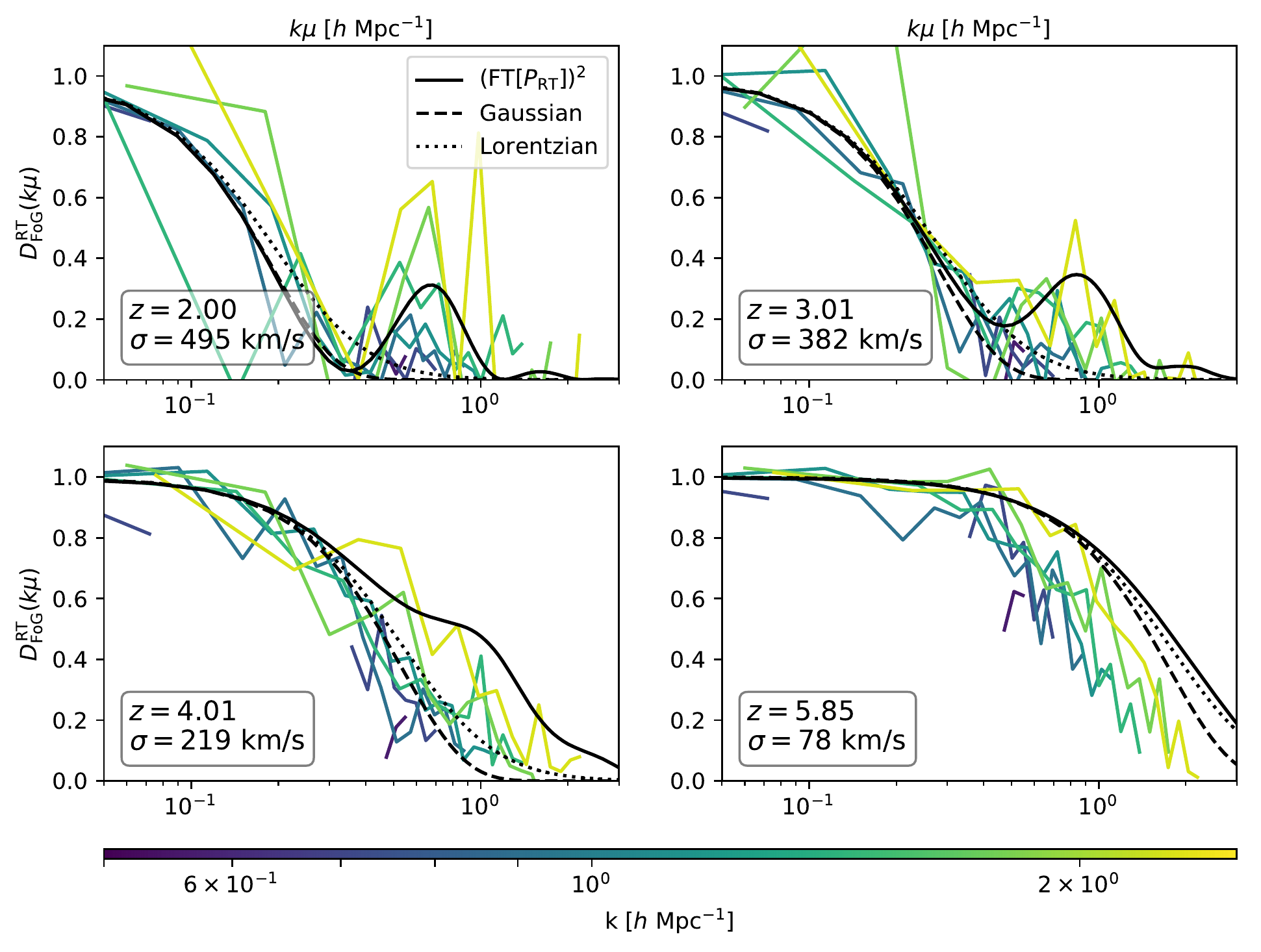}
  \caption{
  Damping factor as a function of the line-of-sight frequency $k_\parallel=k\mu$
  due to Lyman-alpha radiative transfer velocity $v_{\rm RT,global}$. 
  Emitter positions are assumed to coincide with the global peak. 
  Different colored lines represent measurements of $D(k_\parallel)$ from the
  mock catalogs at different total frequencies $k$.
  The black solid lines show the Fourier transform of the 1-point PDF (see Eq.~\eqref{eq:dampingfactor}), while the dashed and dotted lines show the Gaussian
  and the Lorentzian forms respectively (see Eq.~\eqref{eq:DFoG_Gaussian}/\eqref{eq:DFoG_Lorentzian}).
  \textbf{Left-to-right, top-to-bottom}: z=2.00, z=3.01, z=4.01, z=5.85}
\label{fig:PSdamping_redshift_vrt}
\end{figure*}

\begin{figure*}
\centering
  \includegraphics[width=1.0\linewidth]{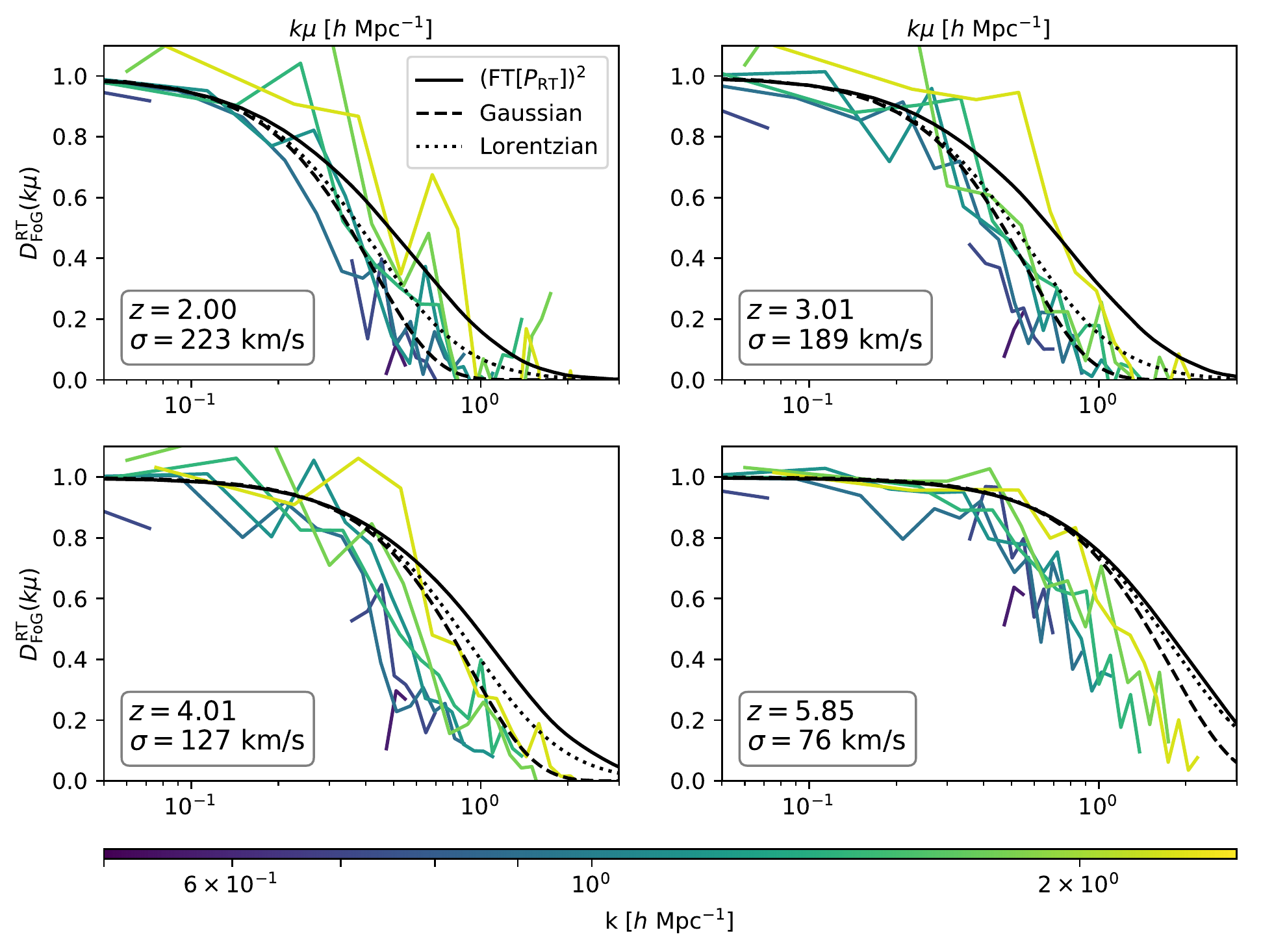}
  \caption{Damping factor as a function of the line-of-sight frequency
    $k_\parallel=k\mu$ due to Lyman-alpha radiative
	  transfer velocity $v_{\rm RT,red}$. Emitter positions are assumed to coincide with the global
  	  peak in the red wing with respect to the halo's frame.
  	  Different colored lines represent measurements of $D^{\rm RT}_{\rm FoG}(k_\parallel)$
      from the mock catalogs at different total frequencies $k$.
  The black lines show the Fourier transform of the 1-point PDF (see Eq.~\eqref{eq:dampingfactor}), the Gaussian
  and the Lorentzian form (see Eq.~\eqref{eq:DFoG_Gaussian}/\eqref{eq:DFoG_Lorentzian}).
  	\textbf{Left-to-right, top-to-bottom}: z=2.00, z=3.01, z=4.01, z=5.85}
\label{fig:PSdamping_redshift_vrt_RED}
\end{figure*}

In Figures~\ref{fig:PSdamping_redshift_vrt} and~\ref{fig:PSdamping_redshift_vrt_RED}, we show such direct comparisons for the global-peak and the red-peak cases, respectively. 
We show the measurements of the ratio of the two redshift-space spectra as a function of the mode parallel to the line-of-sight, $k_\parallel=k\mu$, color-coding them by their absolute wavenumber, $k$. 
In addition, we plot the expected damping from the velocity PDF as implied by Eq.~(\ref{eq:dampingfactor}) (solid lines) and the two generic fitting functions with the second central moment $\sigma$ of the according velocity PDF (dashed and dotted lines for the Gaussian and the Lorentzian functions, respectively). 
In general, we find a strong damping even on relatively large scales, $k\gtrsim 0.1h/$Mpc, and strength of the damping depends on the peak detection algorithm and redshift.
There is a typical redshift evolution with stronger damping at lower redshift, originating from two contributions. 
First, as seen in Figure~\ref{fig:vPDF}, the velocity distribution widens at lower redshifts, translating to a larger damping scale. 
This is mainly because the neutral hydrogen density in CGM for the threshold sample becomes larger at lower redshift \citep[see also ][]{behrens_impact_2017}.  
Secondly, distance and velocity are related via the mapping from real to redshift space as given by Eq.~(\ref{eq:redshiftspace}). 
As the factor of $aH$ roughly scales as $a^{-1/2}$ in the analyzed redshift range, fixed velocity dispersions correspond to larger damping length scales at lower redshifts.

We also find that the measurements are in fair agreement with the squared direct FT of the one-point PDF (solid lines) at $k\mu \lesssim 1h/$Mpc, although we see some discrepancies in detail. We discuss some possibilities to cause the discrepancies in Appendix~\ref{appendix:Discrepant_DFoG}, but do not provide a decisive reason. 
Comparing the solid lines with the dashed and dotted ones, we see that all of them can qualitatively trace the damping feature as a function of scale, although there are slight differences in detail. 
These differences are expected given the fact that the one-point PDFs follow neither Gaussian nor exponential distributions (see Fig.~\ref{sec:results_PDF_vrt}). 
It is apparent that our measurements are too noisy to conclude which model works the best. 
Nevertheless, this result suggest that the dispersion of the one-point PDF, $\sigma$, is a good proxy for the resulting FoG damping due to RT.

Strong evidence that our measurement is consistent with the direct FT of the one-point PDF comes from the oscillatory behavior in the solid lines.
This can be understood as follows. 
Suppose that we model the double-peaked PDF as $P_{\rm RT}(v)=f_{1}(v)+f_{2}(v+\Delta v)$ where $f_{i}(v)$ is a symmetric distribution with a peak and $\Delta v$ denotes the separation of the two peaks. 
Then we have 
\begin{eqnarray}
  \label{eq:doubledamp}
  && D^{\rm toy}_{\rm FoG} =\left|{\rm FT}\left[P_{\rm RT}\right]\right|^2 = \left|{\rm FT}\left[f_1(v)+f_2(v+\Delta v)\right]\right|^2\nonumber\\
  &=& \left|{\rm FT}\left[f_1\right]\right|^2+\left|{\rm FT}\left[f_2\right]\right|^2 
  + 2{\rm FT}\left[f_1\right]{\rm FT}\left[f_2\right]\cos\left[k\Delta v\right]
\end{eqnarray}
From this toy model we see that the first two terms give the FoG-like damping due to each of the single-peaked distributions, while the last term gives an oscillation due to $\Delta v$. 
In other words, the oscillatory behavior originates from the double peak distribution in the global-peak case. 
The oscillations disappear at higher redshifts, since the second peak in the PDF at a blue end is suppressed by the attenuation due to neutral hydrogen in IGM, as we visually confirm in Fig.~(\ref{sec:results_PDF_vrt}).
Furthermore, one can quantify when the oscillatory term becomes dominant by approximating both peaks by a Gaussian distribution with the same dispersion $\sigma$:
\begin{equation}
 D^{\rm toy,Gaussian}_{\rm FoG}  = \underbrace{\exp\left(-\sigma^2k^2\right)}_{D^{\rm Gaussian}_{\rm FoG}}\underbrace{\left[w_1^2+w_2^2+2w_1w_2\cos\left(k\Delta v\right)\right]}_{D^{\rm osci}_{\rm FoG}},
\end{equation}
where $w_1$ and $w_2$ are the relative contribution of the peaks to the PDF such that $w_1+w_2=1$. By looking at the leading, second-order term of $D^{\rm Gaussian}_{\rm FoG}$ and $D^{\rm osci}_{\rm FoG}$ (assuming $k\Delta v \ll 1$ and  $k\sigma \ll 1$), we find that the cosine term can dominate if 
\begin{align}
\label{eq:GaussVScosine}
    \frac{\Delta v}{\sigma} > \frac{1}{\sqrt{w_1 w_2}}.
\end{align}
In addition, notice that the impact of $D^{\rm osci}_{\rm FoG}$ is limited as it has a lower bound:
\begin{align}
\label{eq:Dosci}
    D^{\rm osci}_{\rm FoG}\ge \left(2 w_1 - 1\right)^{2}.
\end{align}
Applying this double-peaked Gaussian to our mock data, we find reasonable fits that could be further improved by substituting the Gaussians with a better fits for the peak PDFs. At all redshifts except for z=5.85, where the blue peak has completely vanished, the Gaussian dominates according to Eq.~\ref{eq:GaussVScosine}. Looking at Fig.~\ref{fig:PSdamping_redshift_vrt}, we nevertheless see that the impact of the oscillatory part is weak. This can be attributed to the strong asymmetry in the height of the two peaks (e.g. $w_1\sim 0.06$ at $z=4.01$), which reduces the amplitude of $D^{\rm osci}_{\rm FoG}$ according to Eq.~\ref{eq:Dosci}.

\section{Discussion}
\label{sec:discussion}

We have shown that significant FoG-like damping arises in the large-scale LAE clustering in redshift space as the peak positions of observed LAE spectra are affected by RT. 
We now discuss the most important steps and caveats concerning our findings for the FoG-like damping factor, $D_{\rm FoG}(k,\mu)$, for LAEs, and also investigate possibilities to mitigate the RT damping.

In Section~\ref{sec:discuss_spectra}, we discuss the shape of spectra arising in our simulations and possible shortcomings in our modeling. Next, we discuss the line-of-sight localization method, which reduces a spectrum to a radiative velocity $v_{\rm RT}$ in Section~\ref{sec:discuss_localization}. 
Afterwards, we seek to use additional information from the spectra to reduce the damping effect in Section~\ref{sec:discuss_correction}. 

\subsection{Spectra}
\label{sec:discuss_spectra}

We obtain individual spectra in our simulations by using an aperture method with a 3 arcseconds radius as introduced in Section~\ref{sec:methods}. 
Using this simple detection algorithm, there are only three relevant parameters impacting the damping factor that are related to instrumental specifications in a real observation: 
The spatial resolution (aperture size), the spectral resolution and the number density threshold. 
The number density threshold roughly corresponds to a flux threshold which should be determined by the signal-to-noise ratio in real observations.
We show the moderate impacts of both spectral resolution and aperture size on the $v_{\rm RT}$ distribution in Appendix~\ref{sec:spectralres} and~\ref{sec:angularres}. Similarly we have already discussed the impact of the number density in Section~\ref{sec:results_PDF_vrt}.\par 

In our simulations, we obtain a manifold of different spectral shapes as shown in Figure~\ref{fig:spectra_selection}. 
Most prominently in this work, we seem to overpredict the amount of double peaked emitters as we mentioned in Section~\ref{sec:results}.
As the original Illustris simulations do not resolve the ISM, we effectively set the post-ISM spectrum to that of the input photons which follow a Gaussian distribution. 
Thus, many of the photons, which are close to the line center, will be reprocessed on the CGM (rather than the ISM) scales to exhibit the rich dataset of spectra we obtain.
It is not trivial to see how the lack of ISM modeling will affect the resulting spectra other than that it most likely reduces the fraction of double peaked profiles when an ISM model with galactic outflows (\citet{BonilhaMonteCarlocalculations1979}) is chosen. In fact, mrore ecent work attributes the frequency redistribution to the ISM scales rather than the CGM scales and includes IGM attenuation as an additional separated step (\cite{InoueSILVERRUSHVIsimulation2018}, \cite{Gurung-LopezLyaemitterscosmological2019}). 
We do not expect that the addition of a sophisticated ISM model will change the qualitative nature of the newly modeled FoG-like damping: 
An ISM model, particularly with outflows, should not introduce additional large-scale correlations, since the large-scale correlation is mainly driven by the IGM attenuation at a blue end of the input spectrum \citep{behrens_impact_2017}. 
Therefore, we expect the qualitative modeling as a FoG-like damping in redshift space remains modeling even when introducing a more sophisticated spectral modeling on ISM scales.

There is a series of other modeling shortcomings in our work including the lack of dust \citep{LaursenLyaRadiativeTransfer2009}, subgrid modeling (for sub-parsec clumps) (\citet{GronkeMirrorsWindowsLymanAlpha2016}, \citet{GronkeResonantlinetransfer2017}) and spatial resolution. 
We only consider Lyman-$\alpha$ emission from recombination in the star-forming regions. Contributions from collisional excitation can be significant and show a bluer spectral signature~\citep{SmithphysicsLymanescape2019}. As initial photons are spawned from a point-source within the LAEs in our simulations and thus not reflecting different physical environments, such spectral modifications cannot be captured within our framework. As we are concerned with individual LAEs' spectra, we can neglect the small, fluorescent contributions in the IGM \citep{Dijkstra:2017pp}.

We note that the original Illustris simulations show a growing excess of neutral hydrogen (a factor of $\sim 3$ between $z=2$ and $4$) compared to observations~\citep{Diemer2019} on galaxy scales where frequency diffusion of Lyman-$\alpha$ photons gives rise to the FoG-like damping. This leads to an overestimate of the dispersion in the $v_{\rm RT}$-PDF. For the Neufeld solution \citep{Neufeldtransferresonancelineradiation1990} such an hydrogen excess would correspond to an overestimate in the peak offsets $v_{\rm RT}$ of roughly $31\%$ ($\Delta v \propto \tau^{1/3} \propto n_{\rm HI}^{1/3}$).

As found in \citet[][]{behrens_impact_2017} for the real-space clustering, large-scale correlations and LAE mock observables significantly change with the resolution of the underlying neutral hydrogen distribution in the RT simulations. Similarly, we give a possible explanation for the lack of detection of the additional redshift space distortion in \cite{Zheng2011} due to its limited hydrodynamic resolution in Section~\ref{sec:comparestudies}. 

Overall, we thus expect the observational spectral shapes to differ from those found in our simulations. 
As a result, our simulations do not reproduce the observed Lyman-$\alpha$ luminosity function as addressed in \citet[][]{behrens_impact_2017}.
This problem of reproducing observables is common when not explicitly calibrating against Lyman-$\alpha$ specific observations such as the luminosity function (e.g. calibration against other redshifts, see \cite{KakiichiLyaemittinggalaxiesprobe2016}, \cite{InoueSILVERRUSHVIsimulation2018}).

From the spectra themselves, one can obtain stacked surface brightness profiles $I_{\rm stacked}$ as shown in Figure~\ref{fig:vPDF_stacked}.
However, it is important to realize that it is the \textit{unobservable} one-point velocity PDF, $P_{\rm RT}$, that mainly determines the additional FoG-like damping along the line-of-sight direction.
Since the relation between $I_{\rm stacked}$ and $P_{\rm RT}$ is non-trivial, there is no \textit{a priori} way of determining the damping factor from the stacked profiles.
Nevertheless, we find an empirical relationship between the square root of the second central moment of $I_{\rm stacked}$ and $P_{\rm RT}$: 
\begin{align}
  \label{eq:sigmarelation}
\sigma_\mathrm{PDF}\approx0.4\cdot\sigma_\mathrm{stacked}
\end{align}
for $n_{\rm LAE}=10^{-2}$ h$^{3}$Mpc$^{-3}$ at $z=3$. 
This relation typically changes less than $10\%$ among redshifts from 3.01 to 5.85. 
At redshift $z=2.0$ this proportionality factor is consistenly higher
by a factor of $\sim 25\%$. 
The number density has a large impact on the resulting relation and rising from $0.4$ to $0.55$ when restricting the number density from $0.01$
to $0.001$ h$^{3}$Mpc$^{-3}$. 
This change is caused by an increased dispersion in the velocity distribution, while the width of the stacked profiles is nearly
constant independent of imposed number density threshold. 
This relation was found in the halos' rest frame, which is unknown unless a secondary emission line (e.g., H$\alpha$) is measured and its radial position is identified.\par

\subsection{Localization along the Line-of-Sight}
\label{sec:discuss_localization}
We proposed two simple localization methods to identify LAE's radial position in redshift space for given  Lyman-$\alpha$ spectra: Either by identifying the position of the maximal spectral flux with the LAE's position (referred to as `the global peak'), or by identifying the LAE position with the maximal spectral fluxes at wavelengths longward of the actual LAEs' radial positions as used for the $v_{\rm RT,red}$ distribution (referred to as `the red peak only'). 
Combination of our simulated spectra with the choice of the localization methods gives us the one-point velocity PDF, $P(v)$, which is a key quantity to understand the resulting FoG damping. 
Nevertheless, let us briefly discuss potential issues in $P(v)$ in light of the two localization methods here. 

The major issue with the global-peak method is closely related to the fact that we tend to overestimate the number of double peaked spectra in our simulations. 
As a consequence, the velocity distribution $P_{\rm RT}(v)$ has a strong double peak feature as well, as we see in Figure~\ref{fig:vPDF}, which does not most likely represent real observations.
Furthermore, as shown in Section~\ref{sec:damping}, such a double peaked velocity distribution induces a larger FoG damping as compared with the $v_{\rm RT,red}$ distribution. 
In a simple toy model for a double peaked velocity distribution, we see that two additional terms occur strengthening the damping, where one is induced by the second peak's width and the second by the separation between the peaks.

The second detection method does not suffer from these additional damping contributions and can be applied to both the double-peaked spectra and single-peaked spectra.
However, there remains a chance to get an oscillatory damping component as well in this localization method if some LAEs' visible peaks are blue of the line center, which is not distinguishable from a red peak in real observations without a complementary emission line. 
Nevertheless, it is ad hoc to completely ignore the blue part in the red-peak only method without any physically reasonable reasons. 
We thus stress that the velocity PDFs with two methods and the resultant FoG damping strength are showcases of two extreme situations.

The localization methods so far only use a single peak's frequency shift for the determination of the line-of-sight position. 
However, we can in principle try to use other information available through the individual spectra to determine the line-of-sight position. 
In the coming section, we investigate methods to further reduce the radiative transfer damping in the clustering signal based solely on additional information of the features in the Lyman-$\alpha$ spectrum.

\subsection{Correction to mitigate the impact of RT}
\label{sec:discuss_correction}

\begin{figure}
\centering
  \includegraphics[width=1.0\linewidth]{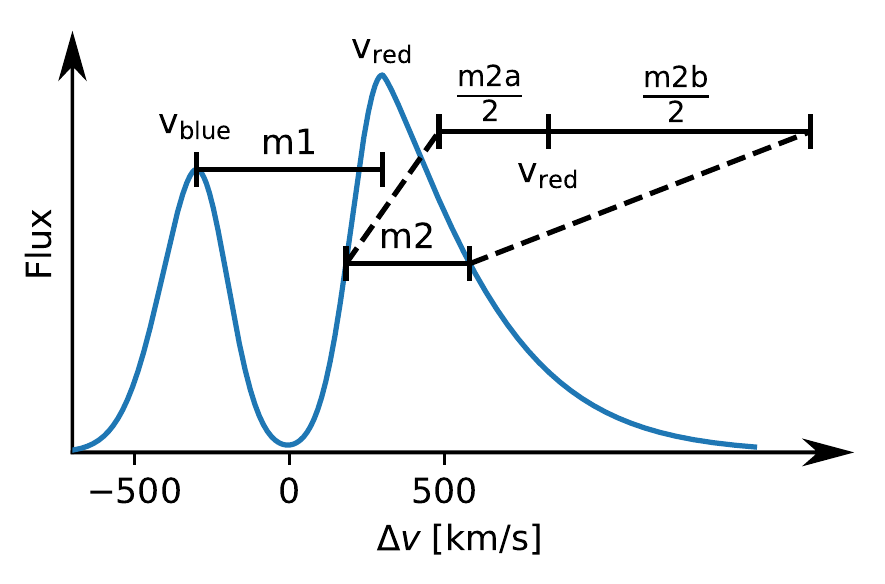}
  \caption{Sketch of the proposed correction methods (see text) applied to a example double peak spectrum.}
\label{fig:correctionsketch}
\end{figure}

\begin{figure}
\centering
  \includegraphics[width=1.0\linewidth]{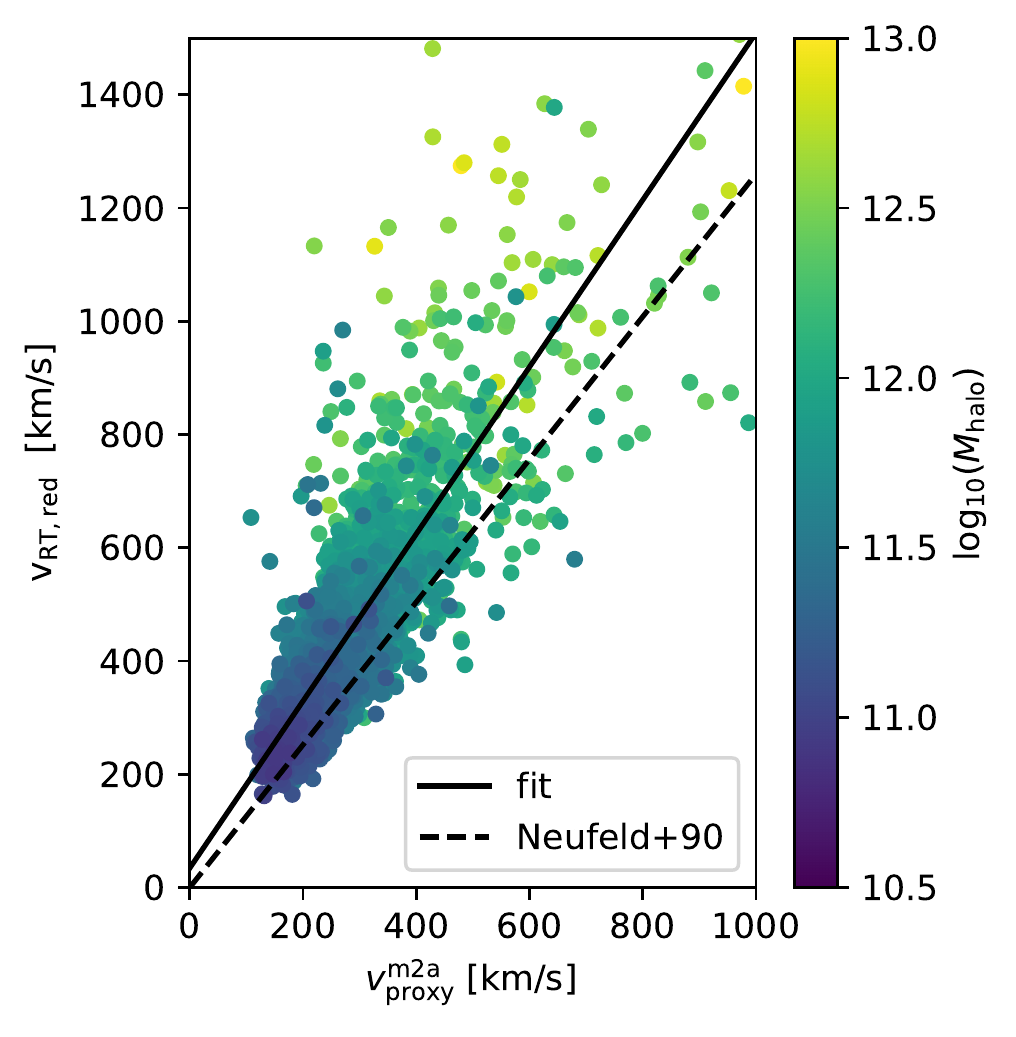}
  \caption{Scatter plot of $v_{\rm RT, red}$ and $v_{\rm proxy}^{\rm m2a}$ for emitters
	  detected at number density threshold $n_{\rm LAE}=10^{-2}$ h$^{3}$Mpc$^{-3}$
  and redshift $z=3.01$. We also show a linear fit (see Eq.~\eqref{eq:linregress}),
  the slope expected from an optically thick spherical HI distribution
  \citep{Neufeldtransferresonancelineradiation1990}.}
\label{fig:vRTcorrection_z3.01_m2a}
\end{figure}

We can try to correct for the presented Lyman-$\alpha$ radiative transfer distortion effect by utilizing the full spectral information available for the emission line. 
To do so, we investigate correlations of the peak offset to other
characteristics of the spectra. 
These include: Half the separation between red and blue peak (\textbf{m1}), the full width at the half maximum (FWHM) of the red peak (\textbf{m2}), twice the half width at the half maximum (HWHM) of the red peak towards the line center (\textbf{m2a}) and twice the HWHM of the red peak away from the line center (\textbf{m2b}). Hence, \textbf{m2} is the average of \textbf{m2a} and \textbf{m2b}. A visualization of these methods is shown in Figure~\ref{fig:correctionsketch}.

We choose to divide the second method (\textbf{m2}) into \textbf{m2a} and \textbf{m2b} as we expect different physical causes for their respective wing shape. The wing towards the line center should be strongly influenced by the IGM attenuation. This has been shown to hold true even at low redshifts very close to the line center~\citep{laursen_intergalactic_2011}. 
However, the IGM is transparent for the wing away from the line center and the red peak, and therefore should only be impacted by small-scale frequency diffusion. 
Indeed we will find large differences in the results from \textbf{m2a} and \textbf{m2b}.

We calculate slope $f$ and offset $v_\mathrm{offset}$ for a linear regression of form 
\begin{align}
\label{eq:linregress}
  v_\mathrm{predict} = f\cdot v_\mathrm{proxy} + v_\mathrm{offset}
\end{align}
and also the corresponding Pearson correlation coefficient $p$. 
$v_\mathrm{proxy}$ denotes the respective velocities by \textbf{m1}, \text{m2}, \textbf{m2a} and \textbf{m2b}. $v_\mathrm{predict}$ is the predicted correction for the corresponding peak position $v_\mathrm{RT}$.
With such prediction, we can correct the former $v_\mathrm{RT}$ distribution as

\begin{align}
\label{eq:correction}
v_\mathrm{RT,corr} = v_\mathrm{RT}-v_\mathrm{predict}.
\end{align}

When a blue peak is additionally available for \textbf{m1}, we use the detection algorithm introduced in Section~\ref{sec:isotropic}: 
Peaks are identified as connected areas in $F_\lambda(\Delta\lambda)$ above such a threshold value that a given number density $n_{\rm LAE}$ is reached. 
Additionally, we require that the maximal brightness of a peak needs to exceed $10\%$ of the maximal brightness of the brightest peak. If only one peak is available the emitter is excluded when computing $m_1$.

The linear regression in Eq.~(\ref{eq:correction}) is motivated by Neufeld's solution for which both peaks' FWHM (i.e. \textbf{m2}) and peak offset scale with $\tau^{1/3}$ for an optically thick spherical HI distribution. 
From Neufeld's derivation, a slope of $1.26$ is derived for the relation between offset and FWHM \citep{Neufeldtransferresonancelineradiation1990}.
We expect that anisotropies, dust, the velocity field and IGM interaction introduce a significant scatter as well as noticeable change to the slope parameter. Such a correlation has been found in observations \citep[see e.g.,][]{VerhammeRecoveringsystemicredshift2018}.

In general, we find \textbf{m1} to perform the best with a correlation coefficient of $p\gtrsim 0.95$ across the studied redshift range. 
However, most of observed LAEs are not doublely peaked, thus only methods \textbf{m2}, \textbf{m2a}, \textbf{m2b} are available. 
For these, we find \textbf{m2a} with $0.82 \lesssim p\lesssim 0.85$ to perform the best. 
\textbf{m2} performs slightly worse with $0.77 \lesssim p\lesssim 0.78$ and \textbf{m2b} significantly worse with $0.40 \lesssim p\lesssim 0.65$. 
Note that we restrict the emitter sample for the regression to those with $v_{\rm RT}<800\ $km/s in order to allow comparisons to observational studies and hinder the most massive emitters to dominate the fit due to their increased scatter.

An example of the correlation between $v_{\rm RT}$ and $v_{\rm proxy}$ is shown in Figure~\ref{fig:vRTcorrection_z3.01_m2a} for \textbf{m2a} at $z=3.01$ and a
number density threshold of $0.01$~Mpc$^{-3}$h$^{3}$. 
The best fit in this case yields $v_\mathrm{predict} = 1.47\cdot v^{\rm m2a}_\mathrm{proxy} + 37\mathrm{~km/s}$ with
a Pearson coefficient of $0.83$.
For other redshifts, we find a slope of $1.39$, $1.43$, $1.70$ ($z=2.0$, $z=4.01$, $z=5.85$) with similar or lower offsets. 
Note that the constant offset itself is irrelevant for the damping scale as it does not change the pairwise velocity distribution. 
We find that there is a strong dependence of the slope parameter on the chosen number density threshold.
The slope parameter increases as the number density is decreased.
This finding is related to the dependence of host halo mass: As shown in Figure~\ref{fig:vRTcorrection_z3.01_m2a}, points with larger halo mass tend to have a slightly steeper slope with larger scatter.  

For \textbf{m1}, we find that the slope is close to unity and in very good agreement with observations \citep{VerhammeRecoveringsystemicredshift2018}, while for methods using the peaks' width (\textbf{m2},\textbf{m2a},\textbf{m2b}) we find a slope above unity that makes it slighly higher than the Neufeld solution and considerably higher compared to observations and shell models (\citet{ZhengAnisotropicLymanalphaEmission2014}, \cite{VerhammeRecoveringsystemicredshift2018}). 
In addition, we find a significant dependence on selection criteria and redshift, possibly explaining some of the discrepancies between our results and the literature apart from the mentioned modeling shortcomings in our simulations. 

\begin{figure}
\centering
  \includegraphics[width=1.0\linewidth]{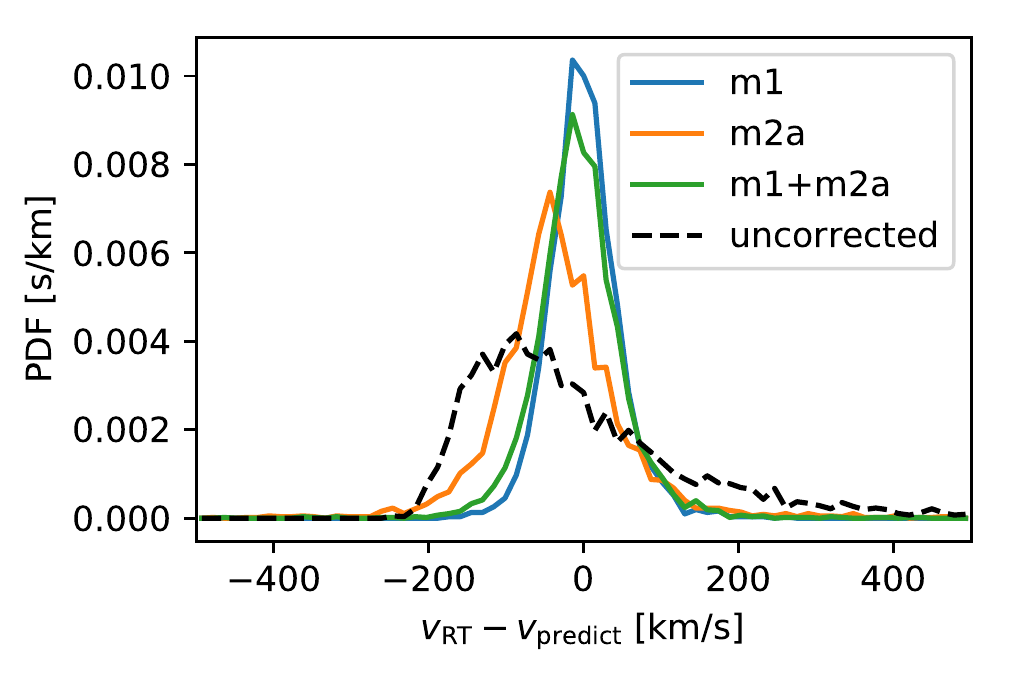}
  \caption{Radiative velocity distribution after different correction schemes
    at a number density threshold of $n_{\rm LAE}=10^{-2}$ h$^{3}$Mpc$^{-3}$ and redshift
    $z=3.01$. The mean velocity is subtracted from each distribution. The dashed line shows the uncorrected distribution.
    The dispersion for the three correction schemes reduce to $48\ $km/s, $92\ $kms/s, $65\ $km/s (m1, m2b, m1+m2a) compared to $189\ $km/s when not corrected.}
\label{fig:vRTcorrection_z3.01_PDF}
\end{figure}

Similarly to~\cite{VerhammeRecoveringsystemicredshift2018}, we can correct for the systemic redshift offset by subtracting the FWHM. 
Doing so consistently reduces the velocity dispersion by a factor of 2-3 across the simulated redshift range and accordingly shortens the damping scale.
Figure~\ref{fig:vRTcorrection_z3.01_PDF} shows the corrected distribution according to Eq.~(\ref{eq:correction}) at the same redshift and number density threshold. 
Also the third and fourth central moment are reduced significantly: the normalized moments (skewness, kurtosis) shrink such that a Gaussian fit becomes feasible.

\section{Conclusions}
\label{sec:conclusions}

In this paper, we have studied the clustering of LAEs in redshift space using a full RT simulation on Illustris.
We find a new kind of the RSD effect due to RT, and our executive summary is the following:

\begin{itemize}
    \item The additional redshift space distortion stems from small-scale frequency diffusion of the Lyman-$\alpha$ line leading to a shift of the spectral peaks. 
    The peak shifts can be larger than those of the peculiar velocity field and thus impact the redshift space clustering signal on larger spatial scales.
    \item We show that the peak shifts from Lyman-$\alpha$ RT damp the power spectrum along the line of sight at scales of $k_{\parallel}\gtrsim 0.1h/$Mpc. 
    We also show that the shifts are mostly independent of the local density and velocity field such that an independent modeling of this shift's impact can be done. This is similar to that of the Fingers-of-God effect due to random motion of galaxies, i.e., in terms of the one-point velocity PDF.
    However, the functional form of the damping can be more complex and even involve oscillations due to the double peak nature. 
    \item The strength of the damping depends strongly on the chosen localization method of the Lyman-$\alpha$ emitters in its spectrum. 
    We attempt two extreme scenarios where we find a peak from the entire spectrum (global peak) and only from the spectrum at red end (red peak).
    \item We show that we can mitigate the impact of the distortion by applying a correction scheme of which we present two classes: In the case of double peaked spectra, the midpoint between the two peaks is an excellent proxy for the emitter position. 
    If only a single peak is present, the half-width-half-maximum on the wing towards the Lyman-$\alpha$ line center can be used as a mediocre proxy.
\end{itemize}

We do not attempt to quantify the exact amplitude of the RT effect on actual BAO and RSD galaxy surveys such as HETDEX for the following reasons. 
First, as we often mention, our simulated LAEs do not well reproduce a variety of observables such as the luminosity function likely due to unresolved ISM physics in Illustris.  
Second, the actual impact should depend on resolving power of a spectrum. 
For example, HETDEX has a spectral resolution roughly corresponding to $\Delta v\sim 500$km/s where the resultant PDF is largely smeared out. 
Nevertheless, we stress that the frequency shift can be larger than $\Delta v\sim 500$km/s as seen in Figure~\ref{fig:vRT_vs_vpeculiar_z3.01} and hence we expect that the FoG damping due to RT exists to some extent. 
We leave detailed assessment for future work. 

Although we focus on the clustering of LAEs, it would be straightforward to extend our analysis to the intensity mapping. 
In fact, we visually confirm strongly elongated feature in the Lyman-$\alpha$ intensity map in Figure~\ref{fig:IMcube}: 
For intensity mapping we expect a similar damping to that of the LAEs from their positional offset, but additionally from the width of the spectrum itself which further strengthens the damping.
Observationally, \citet[][]{CroftLargeScaleClusteringLyman2016} reported the large-scale elongation along the line of sight in the cross correlation between the quasars and the Lyman-$\alpha$ intensity map at $z\sim 2$ in the Sloan Digital Sky Survey. 
Even though their more recent study argues that it is due to special environment around quasars given the lack of the cross-correlation signal between the intensity map and the Lyman-$\alpha$ forest \citep[][]{CroftIntensitymappingSDSS2018}, the elongation might partly be due to the RT FoG effect. 

As a concluding remark, we give the following general suggestions for cosmological LAE surveys:
\begin{itemize}
    \item We find an empirical relationship between the second central moment for the red and blue peaks in stacked spectra in the halo's rest frame and the dispersion of those peaks in the one-point velocity distribution. Hence, the possible damping can be estimated by measuring the second central moments in the stacked spectra. Stacking these in the halo's rest frame requires knowledge of a secondary emission line tracing the kinematics of the LAE.
    \item Having such second emission line as tracer of the LAEs' velocity for a subset of emitters allows to calibrate the linear fit parameters present in the correction methods based on the Lyman-$\alpha$ spectral features to mitigate the distortion effect.
\end{itemize}

Finally, concerning other target emission lines such as H$\alpha$ for e.g., Euclid \citep[][]{Laureijs:2011aa} and Wide-Field InfraRed Survey Telescope-Astrophysics Focused Telescope Assets \citep[WFIRST-AFTA,][]{Spergel:2013aa} and [OII] for the Subaru Prime Focus Spectrograph \citep[PFS,][]{Takada:2014sf} and Dark Energy Spectroscopic Instrument \citep[DESI,][]{DESI-Collaboration:2016tg}, it is unlikely that RT makes a notable impact on the BAO and RSD measurements, as the transition must be resonant and have high optical depths in the astrophysical environment. 

\section*{Acknowledgements}
We thank Max Gronke for useful discussions concerning the obtained spectra in our simulations.
We are also grateful to Eiichiro Komatsu for his comments on the manuscript. 
This work was supported in part by JSPS KAKENHI Grant Number JP16J01890, and by World Premier International Research Center Initiative (WPI Initiative), MEXT, Japan.
We acknowledge use of the Python programming language \citep{VanRossum1991}, the
use of the Numpy \citep{VanDerWalt2011}, IPython \citep{Perez2007}, and
Matplotlib \citep{Hunter2007} packages. Our analysis benefited from the use of
the following astrophysical packages: astropy \citep{astropy} and halotools
\citep{HearinHighPrecisionForwardModeling2016}.




\bibliographystyle{mnras}
\bibliography{references}


\appendix

\section{Detection Algorithm Variations}
\subsection{Spectral Resolution}
\label{sec:spectralres}

\begin{figure}
\centering
  \includegraphics[width=1.0\linewidth]{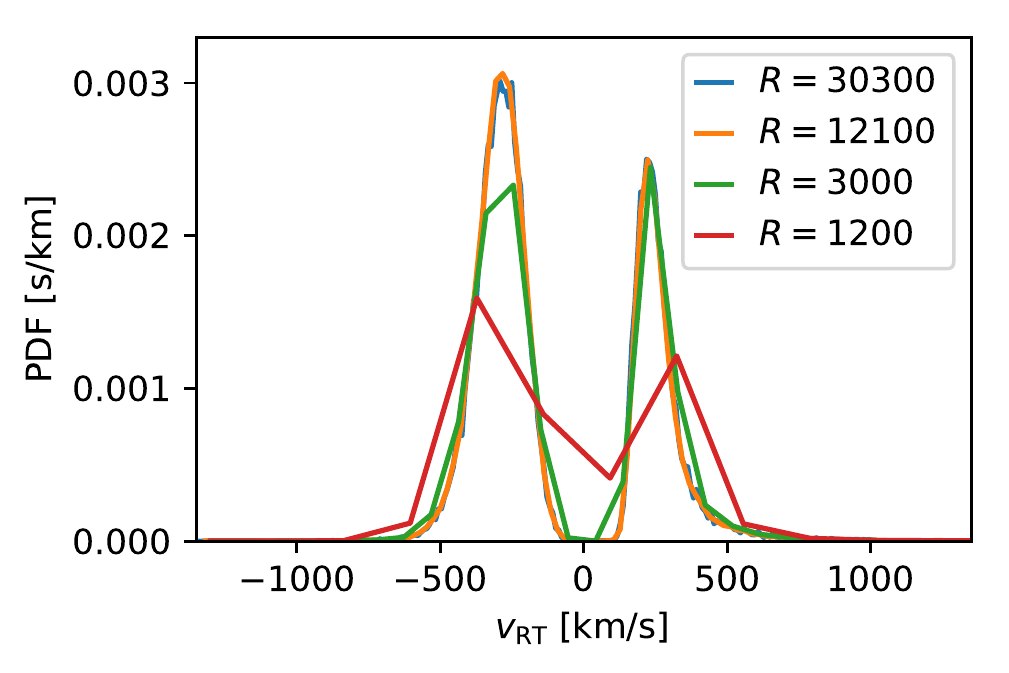}
  \caption{The radiative velocity distribution $v_{\rm RT}$ for varying spectral resolution
    in terms of $R=c/\Delta v$ at z=3.01 for all LAEs (no $n_{\rm LAE}$ restriction).}
\label{fig:vPDF_respower}
\end{figure}

In our study, we do not generally assume a spatial resolution natch to a specific observation both in the spectral and the angular resolutions. 
For the spectral resolution, we infer the redshift space position by adding the line shift to the real space position known from the halo catalogs, while
surveys directly infer the degenerate redshift space position from the line
feature's position. 
Nevertheless, the spectral resolution in simulations and observations should be comparable quantities. 
In our fiducial case, velocities are resolved to $24.7\ $km/s corresponding to a spectral resolution of roughly $R\sim 12000$. 
This exceeds the resolution in HETDEX by more than an order of magnitude ($R\sim 800$, see e.g. \cite{Hill2008}). 
Even dedicated spectroscopic instruments in the search of LAEs such as MUSE only reach values up to $R\sim 3000$ (see e.g.~\cite{BaconMUSE3Dview2015}).

In Figure~\ref{fig:vPDF_respower} we vary the spectral resolution for the
fiducial case at $z=3.01$ and find good reproduction of the characteristics of
the distribution: If existent, the double peaked structure is conserved and the second central moment of individual peaks vary by less than $10\%$.
We have adopted the fiducial resolution of $24.7\ $km/s as we are interested in the physical impact of RT in theory, but simultaneously want to assure sufficient signal to noise in each bin.

\subsection{Aperture Size}
\label{sec:angularres}

\begin{figure}
\centering
\includegraphics[width=1.0\linewidth]{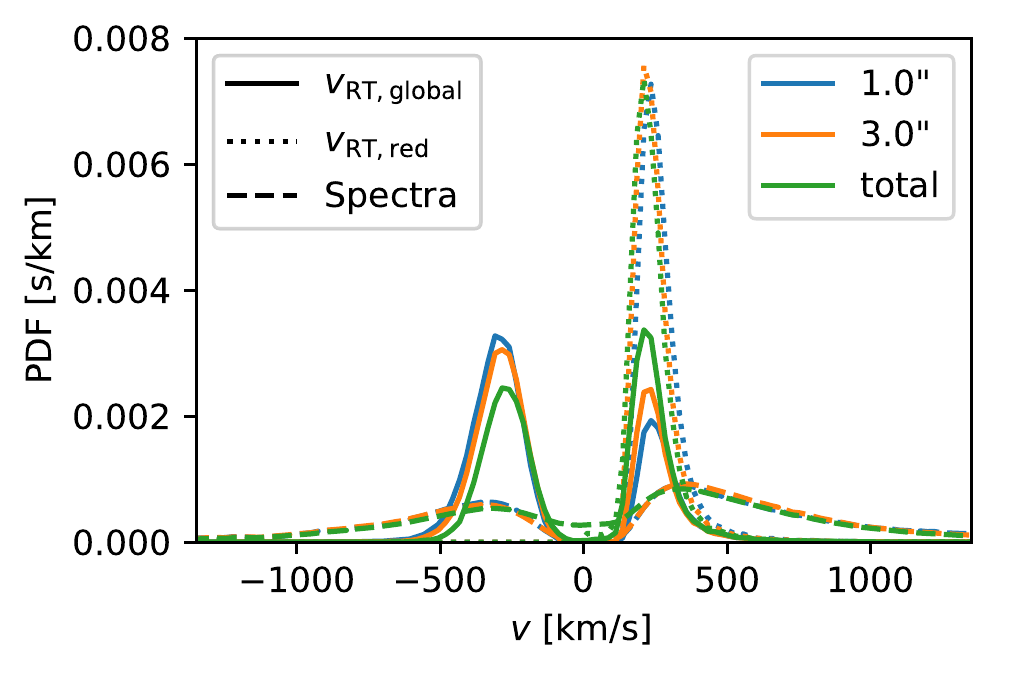}
  \caption{The radiative velocity distributions $v_{\rm RT}$ and stacked spectra
    for varying aperture sizes at $z=3.01$ for all LAEs (no $n_{\rm LAE}$ restriction).}
\label{fig:vPDF_aperturedependency}
\end{figure}

In Figure~\ref{fig:vPDF_aperturedependency}, we show the aperture size
dependency of the v$_{\rm RT}$-probability distribution and of the stacked
profiles at z=$3.01$. The sample includes all simulated LAE rather than
fixing the count to a number density threshold. This reduces the noise, while we
find qualitatively similar results for a restricted sample.

Three qualitative changes occur when increasing the aperture:
A larger aperture appears to favor the red peaks over the blue, the red peak
slightly shifts towards the line center and for large apertures the stacked
spectra show emission in the otherwise deserted trough in the line center.
The second central moments and maxima of the red peaks change by less than
$10\%$, so that the expected change in clustering signal should be similarly small.

\subsection{Refined Detection Algorithm}
\label{sec:isotropic}

\begin{figure}
\centering
  \includegraphics[width=1.0\linewidth]{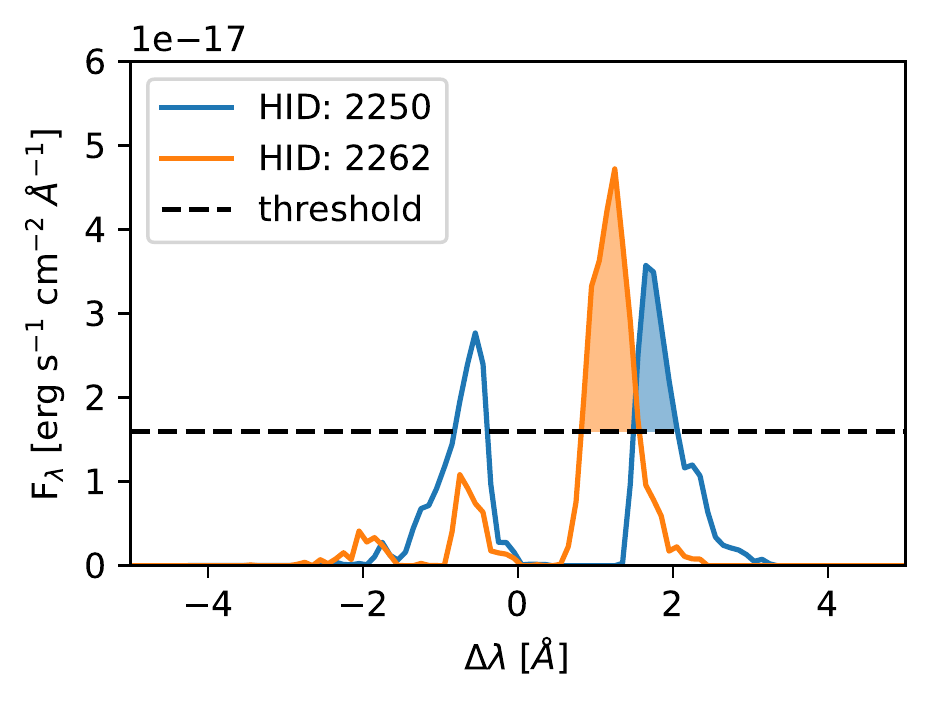}
  \caption{Example of refined detection of two individual emitters with given
    halo ID at z=$3.01$. Dashed line corresponds to the specific flux threshold
    of $1.7\ $erg$\ $s$^{-1}\ $cm$^{-2}\ $\r{A}$^{-1}$ (for number density
    $0.01\ $h$^3$Mpc$^{-3}$) and the shaded region the flux associated with the
    emitters given the refined detection algorithm. For shown emitters, the
    emitter with the highest total flux (which includes all peaks) is different
    from the emitter with the largest flux for the identified peak.}
\label{fig:detection_refined_example}
\end{figure}

In the detection algorithm applied so far, line-of-sight halo positions are inferred
from the spectral peak, while the detectability is determined from the overall
flux. Therefore, the same halo sample is considered as in prior real space
evaluation performed in~\cite{behrens_impact_2017}. However, in redshift space
some emitters' spectra might significantly diffuse while others do not based on
varying small and large scale properties. We therefore expect additional
distortions when emitters are detected by the specific flux. Primarily the
bias is expected to change.

In order to capture the additional distortions, we refined our detection
algorithm as follows: A specific flux threshold t$_{\rm flux}$ is
imposed and only peaks above this threshold are used to identify emitters. Once
a peak (red or global) is identified as the emitter's position, the surface
brightness is integrated around the peak for the spectral range which reaches
t$_{\rm flux}$.

Figure~\ref{fig:detection_refined_example} shows an example of the refined algorithm
for two emitters at z=$3.01$ and how the bias might change: While the spectrum
of HID 2262 has the lower overall flux than HID 2250, it has a larger flux around
the detected peak. Therefore, flux-sorted and number density limited samples
will differ when sorted by total flux or flux associated with the peak.

\begin{figure}
\centering
  \includegraphics[width=1.0\linewidth]{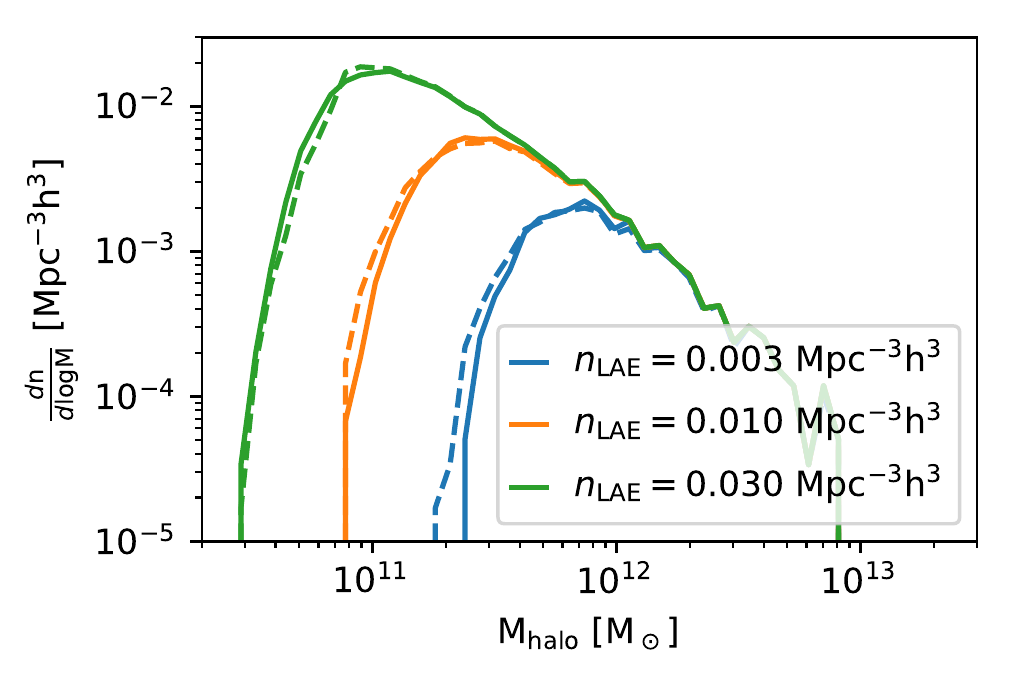}
  \caption{The halo mass function of detected LAEs at $z=3.01$. Solid lines
    represent the fiducial detection method and dashed lines the refined
    algorithm presented. There are only marginal changes in the function at the characteristic mass cut-off imposed by setting the number density threshold $n_{\rm LAE}$.}
\label{fig:HMF_detmode}
\end{figure}

Figure~\ref{fig:HMF_detmode} shows the halo mass function of
detected LAEs for the two discussed detection algorithms. Overall, changes only
occur close to a characteristic mass cutoff that is implied by the number
density threshold and due to the strong correlation between observed flux and
halo mass. As we only find a small change in the mass distribution of detected
LAEs, we do not expect an additional change in bias/isotropic distortion from
radiative transfer in redshift space. We explicitly checked this by calculating
the corresponding velocity distributions and power spectra.

Nevertheless, this detection method is useful in classifying individual peaks
that we will use for identifying blue and red peaks in
Section~\ref{sec:discuss_correction} concerning the correction of the
localization error leading up to the damping. 

\section{Comparison to Prior Studies}
\label{sec:comparestudies}

\begin{figure}
\centering
  \includegraphics[width=1.0\linewidth]{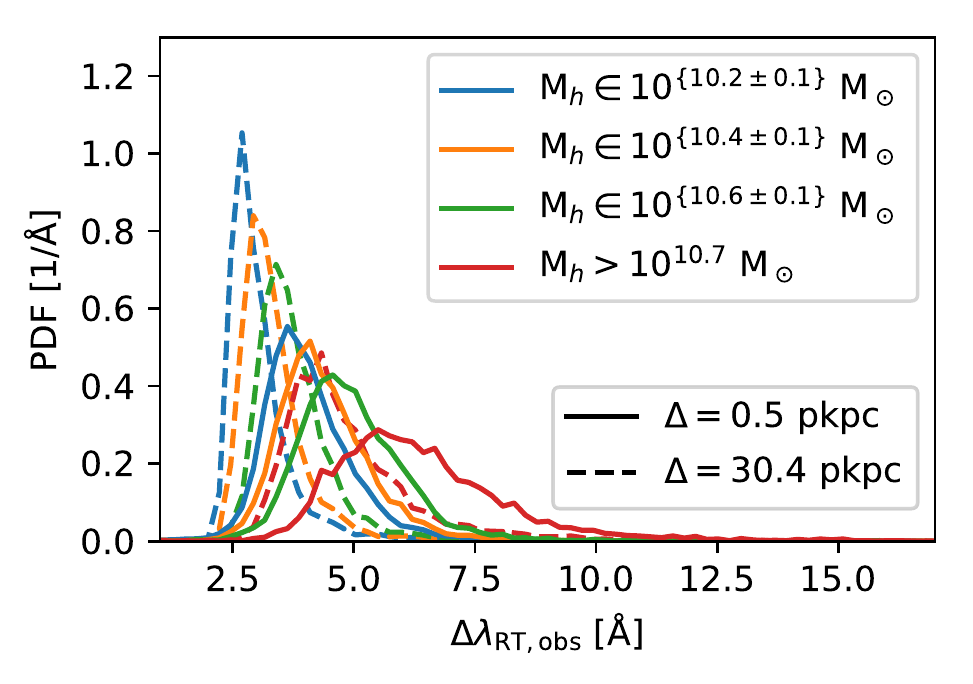}
  \caption{The radiative velocity distribution for varying halo mass range and
  underlying hydrodynamic resolution at redshift $z=5.85$.
  $\Delta\lambda_\mathrm{RT}$ is the radiative velocity as observed wavelength
  shift at $z=0$ with the peculiar velocity $v_{\rm pec}$ having been subtracted
  as usual.}
\label{fig:comparestudies}
\end{figure}

While similar radiative transfer simulations have previously been
performed, the shown redshift space damping has not been observed
by~\cite{Zheng2010}/\cite{Zheng2011}. For consistency, we try to reconcile our
findings with these prior simulations.

As shown, the radiative transfer redshift space distortions significantly
increase at lower redshift: At $z=2.00$, we detect a central second moment
$\sigma$ of the $v_{\rm RT,red}$ distribution of $223\ $km/s, but only a moment
of $76\ $km/s at $z=5.85$. Furthermore, the significance for the redshift space clustering
is reduced at higher redshifts due to the higher Hubble flow as discussed
before.

Additionally, as concluded by~\cite{behrens_impact_2017}, the hydrodynamic
resolution has a large impact on the radiative transfer's results. In
particular, the hydrodynamic resolution largely affects the photon diffusion in
configuration and frequency space determining whether a selection effect is
detected in mock observations or not. Similarly, one might ask whether the hydrodynamic
resolution impacts the radiative transfer redshift space distortion
presented here. We answer this question by comparing the results by analyzing a set of RT
simulations at different resolutions.

Figure~\ref{fig:comparestudies} shows the radiative velocity distribution as a
function of mass at redshift $z=5.85$ of different underlying hydrodynamic
resolution for the radiative transfer post-processing. This plot can directly be compared to the results found in~\cite{Zheng2010} showing a very similar plot (see Figure~6 there).
Given the difference in the baryon modeling following detection method, we find
a very good match between our low resolution run with a grid spacing of
$\Delta=30.4\ $pkpc and the results found by~\cite{Zheng2010} across all mass
bins. Both shown simulations have been imported from~\cite{behrens_impact_2017}
and thus only have an initial photon count of 100 instead of 1000. Thus, minimal
differences to other shown results might be expected.

We find a significant impact of the hydrodynamic resolution
on the velocity distribution. First, the peak of the distribution is shifted
towards larger spectral offsets from the line center at higher resolutions. If
no blue peak is present, as is for $z=5.85$ due to IGM interaction, this overall
shift of the distribution should not affect the damping signal. Second, the
distributions broaden at higher resolutions and thus increases the second central
moment, which enlarges the damping length scale. For the shown resolutions, the
second central moment increases by roughly 20\% across all mass bins at higher
resolution.

\section{Discrepancies in the Damping Factor Modeling}
\label{appendix:Discrepant_DFoG}
As we have shown in Section~\ref{sec:damping}, we have made an attempt to understand the FoG suppression due to RT in terms of the one-point velocity PDF. 
In this appendix, however, we show a hint implying that it is not sufficient to know the one-point PDF to fully model the impact of RT on the redshift-space clustering. 

\begin{figure}
\centering
  \includegraphics[width=1.0\linewidth]{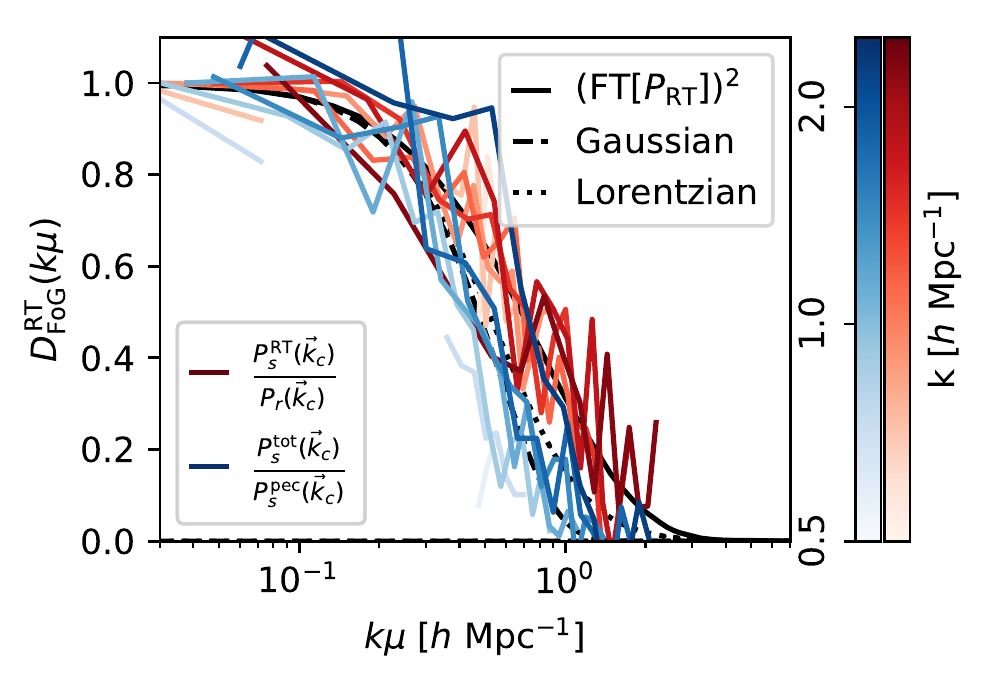}
  \caption{Damping factor as a function of the line-of-sight frequency
    $k_\parallel=k\mu$ due to Lyman-alpha radiative
	  transfer for z=3.01. Emitter positions are assumed to coincide with the red
  	  peak only. Two different ratios $P_s^\mathrm{RT}/P_r$ and
      $P_s^\mathrm{tot}/P_s^\mathrm{pec}$ are shown for the damping factor, see
      text for an explanation. Different colored lines represent measurements of $D(k_\parallel)$
      from the mock catalogs at different total frequencies $k$.}
\label{fig:PSdamping_z3_compare_Dfog_estimators}
\end{figure}

In Figure~\ref{fig:PSdamping_z3_compare_Dfog_estimators}, we present two ratios of the power spectra. 
The blue lines are the ratios of the total redshift-space spectra to ones only with the peculiar velocity component, i.e., $P^{s}_{g,{\rm tot}}/P^{s}_{g,{\rm pec}}$, which are the same results as we showed in Figure.~\ref{fig:PSdamping_redshift_vrt_RED}. 
We also present the ratios of the redshift-space spectra only with the RT velocity component to the real-space power spectra, i.e., $P^{s}_{g,{\rm RT}}/P^{r}_{g}$ as red lines. 
We did not address the latter ratio in the main text, since both the denominator and the numerator are not directly accessible in real observations. 
In Figure.~\ref{fig:PSdamping_redshift_vrt_RED}, it is apparent that the red curves are systematically higher than the blue ones even though both of them are noisy. 

As we stressed in deriving the last equality of Eq.~(\ref{eq:dampingfactor}), a necessary condition which makes the ratio, $P^{s}_{g,{\rm tot}}/P^{s}_{g,{\rm pec}}$, equivalent to the FT of the one-point velocity PDF is following:
\begin{enumerate}
    \item $\langle v_{\rm pec}(x)v_{\rm RT}(x')\rangle=0$, 
    \item $\langle v_{\rm RT}(x)\delta_{g}(x')\rangle=0$, 
    \item and $\langle v_{\rm RT}(x)v_{\rm RT}(x')\rangle=0$.
\end{enumerate}
We argued from Figure~\ref{fig:vRT_vs_vpeculiar_z3.01} that the condition (i) is satisfied at the level of one point, i.e., at $x=x'$. 
Similarly, we did not see any strong evidence which violates the conditions (ii) and (iii) in Figure~\ref{fig:vPDF_pairwise_scaledependency_3in1_z3}. 
In contrast, the condition (i) is not necessary for $P^{s}_{g,{\rm RT}}/P^{r}_{g}$, since it does not involve $v_{\rm pec}$. 
It is encouraging to see that the red lines are more consistent than the blue ones with the direct FT of the one-point PDF (black solid line).
We find this trend holds at other redshift snapshots for both the global-peak and the red-only cases.
This fact implies
that the conditions (ii) and (iii) are basically satisfied. 
We thus may attribute the discrepancy between the red and blue lines to the residual correlation between $ v_{\rm pec}$ and $v_{\rm RT}$, but a conclusive statement with more careful investigation is left for future work. 

\bsp	
\label{lastpage}
\end{document}